\documentclass[10pt,journal,twocolumn,nofonttune]{IEEEtran}

\usepackage{times}
\usepackage{amsmath}
\usepackage{amsthm}
\usepackage{bbm}
\usepackage{bm}
\usepackage{amssymb}
\usepackage{graphicx,dblfloatfix}
\usepackage{cite}
\usepackage{subeqnarray}
\usepackage{xcolor}
\usepackage{multirow}
\usepackage{enumitem}
\usepackage{caption}
\usepackage{tcolorbox}
\usepackage[labelformat=simple]{subcaption}
\usepackage[ruled,vlined]{algorithm2e}
\usepackage{hyperref}
\usepackage{mathtools}
\usepackage{booktabs}
\usepackage{diagbox}
\usepackage{makecell}
\usepackage{threeparttable}
\usepackage{pifont}

\newcommand{\nc}{\newcommand}
\nc{\bsm}{\boldsymbol}
\nc{\mbb}{\mathbb}
\nc{\mbs}{\mathbbmss}
\nc{\mbf}{\mathbf}
\nc{\mcl}{\mathcal}

\newcommand{\cmark}{\ding{51}}%
\newcommand{\xmark}{\ding{55}}%

\DeclareMathOperator*{\argmax}{arg\,max}
\DeclareMathOperator*{\argmin}{arg\,min}
\theoremstyle{definition}

\newtheorem{prop}{Proposition}

\begin{document}

\title{Variance State Propagation for Structured Sparse Bayesian Learning}
%
%
%

\author{Mingchen~Zhang,~
        Xiaojun~Yuan,~\IEEEmembership{Senior~Member,~IEEE},~
        and~Zhen-Qing~He
        \vspace{-0.3cm}
\thanks{M. Zhang, X. Yuan and Z.-Q. He are with the Center for Intelligent Networking and Communications and also with the National Key Laboratory of Science and Technology on Communications, University of Electronic Science and Technology of China, Chengdu 611731, China (e-mail: zhangmingchen@std.uestc.edu.cn; xjyuan@uestc.edu.cn); zhenqinghe@uestc.edu.cn.}
}


\maketitle

\begin{abstract}
We propose a compressed sensing algorithm termed \textit{variance state propagation (VSP)} for block-sparse signals, i.e., sparse signals that have nonzero coefficients occurring in clusters. The VSP algorithm is developed under the Bayesian framework. A hierarchical Gaussian prior is introduced to depict the clustered patterns in the sparse signal. Markov random field (MRF) is introduced to characterize the state of the variances of the Gaussian priors. Such a hierarchical prior has the potential to encourage clustered patterns and suppress isolated coefficients whose patterns are different from their respective neighbors. The core idea of our algorithm is to iteratively update the variances in the prior Gaussian distribution. The message passing technique is employed in the design of the algorithm. For messages that are difficult to calculate, we correspondingly design reasonable methods to achieve approximate calculations. The hyperparameters can be updated within the iteration process. Simulation results demonstrate that the VSP algorithm is able to handle a variety of block-sparse signal recovery tasks and presents a significant advantage over the existing methods.

\end{abstract}


\IEEEpeerreviewmaketitle

\section{Introduction}
In recent years, the compressed sensing (CS) technique, as a new signal acquisition scheme beyond Nyquist sampling, has attracted great interest with a wide range of applications in signal processing \cite{lustig2008compressed,gamper2008compressed} and wireless communications \cite{bajwa2010compressed,berger2009sparse}. Mathematically, given a measurement matrix $\bsm{A} \in \mbb{C}^{M \times N}$ $(M \ll N )$, the goal of CS is to reconstruct an unknown sparse signal $\bsm{x} \in \mbb{C}^{N}$ from the noise-corrupted linear measurements
\begin{equation}
    \boldsymbol{y}=\boldsymbol{A} \bsm{x} + \boldsymbol{w}
\end{equation}
where $\boldsymbol{w} \in \mbb{C}^{M} $ is an additive noise. This problem has been well studied and a variety of algorithms that with guaranteed recovery performance have been proposed, including orthogonal matching pursuit (OMP) \cite{tropp2007signal}, compressive sampling matching pursuit (CoSaMP) \cite{needell2009cosamp}, basis pursuit method \cite{chen2001atomic}, and sparse Bayesian learning (SBL) methods \cite{tipping2001sparse}.

Since the advent of compressed sensing, much research attention has been paid to the reconstruction of a type of sparse signals with additional structure, i.e., sparse signals with the nonzero entries appearing in clusters, namely, \textit{block-sparse} signals. Block-sparsity arises naturally in a variety of practical signals. For instance, in wireless communications, due to the effect of limited local scattering in the propagation environment, the massive multi-input multi-output (MIMO) channel in the virtual angular domain exhibits a block sparsity  \cite{liu2016exploiting,han2017compressed}. In a video surveillance system, foreground moving objects usually occupy a small portion of the camera view, leading to natural block sparsity. Block-sparse patterns also appear in the multiple measurement vector (MMV) problem that deals with the recovery of multiple sparse signal vectors sharing a common nonzero support \cite{eldar2009robust,kim2012compressive}.

For the reconstruction of block-sparse signals, algorithms making an explicit use of the additional block-sparse pattern can yield provably better reconstruction performance than the conventional CS algorithms in which the signals for recovery are assumed to be randomly sparse. A number of CS algorithms are specifically designed for the recovery of block-sparse signals, e.g., Block-OMP \cite{eldar2010block}, mixed $\ell^2/\ell^1$ norm-minimization \cite{eldar2009robust}, and group LASSO \cite{yuan2006model}. These algorithms require strong prior knowledge of the block structure, such as the locations and the lengths of the associated blocks, which are often unavailable in practical applications. Algorithms for structure-agnostic block-sparse signal recovery are also developed. For example, in \cite{yu2012bayesian}, a hierarchical Bayesian Bernoulli-Gaussian prior model was adopted to model both the sparse prior and the
cluster prior, and a Markov chain Monte Carlo (MCMC) sampling method is employed in the inference; in \cite{peleg2012exploiting}, the block-sparse pattern is modeled by a Boltzman machine, and a greedy method was used to simplify the maximum \textit{a posteriori} probability (MAP) estimator; in \cite{zhang2013extension}, the components of the signal are modeled by a number of overlapping blocks, and an expanded block sparse Bayesian learning (EBSBL) was proposed to adaptively exploit intra-block correlation; in \cite{fang2014pattern}, a pattern-coupled hierarchical Gaussian framework was proposed to encourage block-sparse patterns, where the sparsity of each coefficient is controlled by the linear combination of the hyperparameters of itself and its neighbors.
Although they require little or even do not require the prior information of the sparse patterns, these algorithms typically perform far away from the genie bound in which the location of nonzeros is known \textit{a priori}.

Due to its outstanding performance and low complexity, the message passing technique has been employed in the design of CS algorithms for a decade. Representative message passing based CS algorithms include approximate message passing (AMP) \cite{donoho2009message}, generalized approximate message passing (GAMP) \cite{rangan2011generalized}, expectation-maximization Gaussian-mixture approximate message passing (EM-GM-AMP) \cite{vila2013expectation}, and turbo compressed sensing (Turbo-CS) \cite{ma2014turbo}. Message passing based CS algorithms for block-sparse signals have also been proposed in \cite{chen2017structured,chena2019structured,he2019super}. In \cite{chen2017structured}, structured turbo compressed sensing (STCS) was developed for massive MIMO channel estimation. By combining a Markov prior into the Turbo-CS framework, STCS fully utilizes the knowledge of block sparsity and shows superior recovery performance. It is known that message passing based CS algorithms are sensitive to the choice of measurement matrices, since the sum-product rule used in message calculation requires the independence of relevant messages. The convergence of the AMP is guaranteed when the elements of the measurement matrix $\boldsymbol{A}$ are independently and identically distributed Gaussian and the length of the signal $N$ is large \cite{donoho2009message}. Turbo-CS relaxes the requirement on $\boldsymbol{A}$ such that $\boldsymbol{A}$ is allowed to consist of rows randomly selected from an orthogonal basis \cite{chen2017structured} or is right-rotationally invariant \cite{ma2014turbo}. However, the performance of the message passing based algorithms may deteriorate severely when other measurement matrices are involved. It is therefore desirable to design a message passing based CS algorithm that is able to handle a wider range of applications.

In this paper, we propose a new message passing based CS algorithm for the reconstruction of block-sparse signals.
A novel hierarchical Gaussian framework is deployed to model the sparse prior, in which the unknown signal components are modeled as independent Gaussian variables with zero mean and certain variances. Each variance is still regarded as a random variable and assigned a Bernoulli-Gamma prior with a support indicator. These support indicators, a.k.a. the \textit{state variables}, are assigned as a Markov random field (MRF) to capture the block sparsity. 
Such a prior has the potential to encourage block-sparse patterns and suppress ``isolated coefficients'' whose pattern is different from that of its neighboring coefficients. Message passing is performed based on the hierarchical probability model and an iterative algorithm is accordingly developed to estimate the block-sparse signal. For messages difficult to compute, we give approximate calculation methods. The model hyperparameters are updated during the iteration.
Since the state of the variances plays a crucial role in message propagation, we refer to our proposed algorithm as \textit{variance state propagation} (VSP).
Our numerical results show that VSP inherits the superior performance of the message passing based compressed sensing algorithms while maintaining the robustness to the choice of measurement matrices.

The rest of the paper is organized as follows. In Section \ref{ProbModel}, we introduce the MRF-combined hierarchical probability model that characterizes the sparse prior and the pattern dependencies among the signal components. An iterative message passing algorithm is developed in Section \ref{SecVSP} to estimate the block-sparse signal. Section \ref{SecVSP} contains approximate methods for messages that are difficult to compute, and learning methods for model hyperparameters. Simulation results are provided in Section \ref{Simu}, followed by concluding remarks in Section \ref{Conclusion}.

\begin{figure*}[t]
    \centering
    \begin{subfigure}{0.49\textwidth}
    \centering
    \includegraphics[height=5cm]{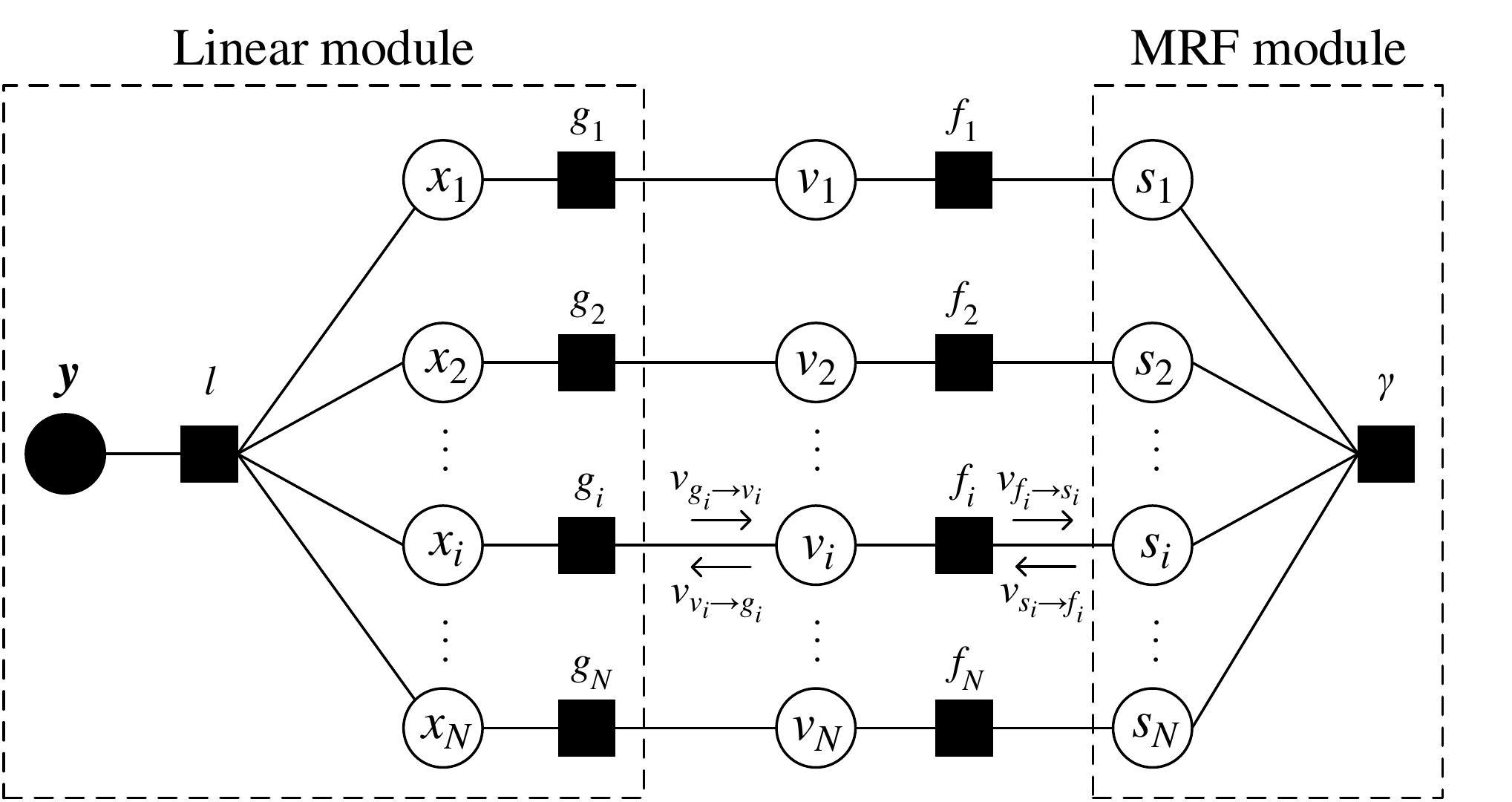}
    \caption{}
    \label{fg2a}
    \end{subfigure}
    \hfill
    \begin{subfigure}{0.49\textwidth}
    \centering
    \includegraphics[height=5cm]{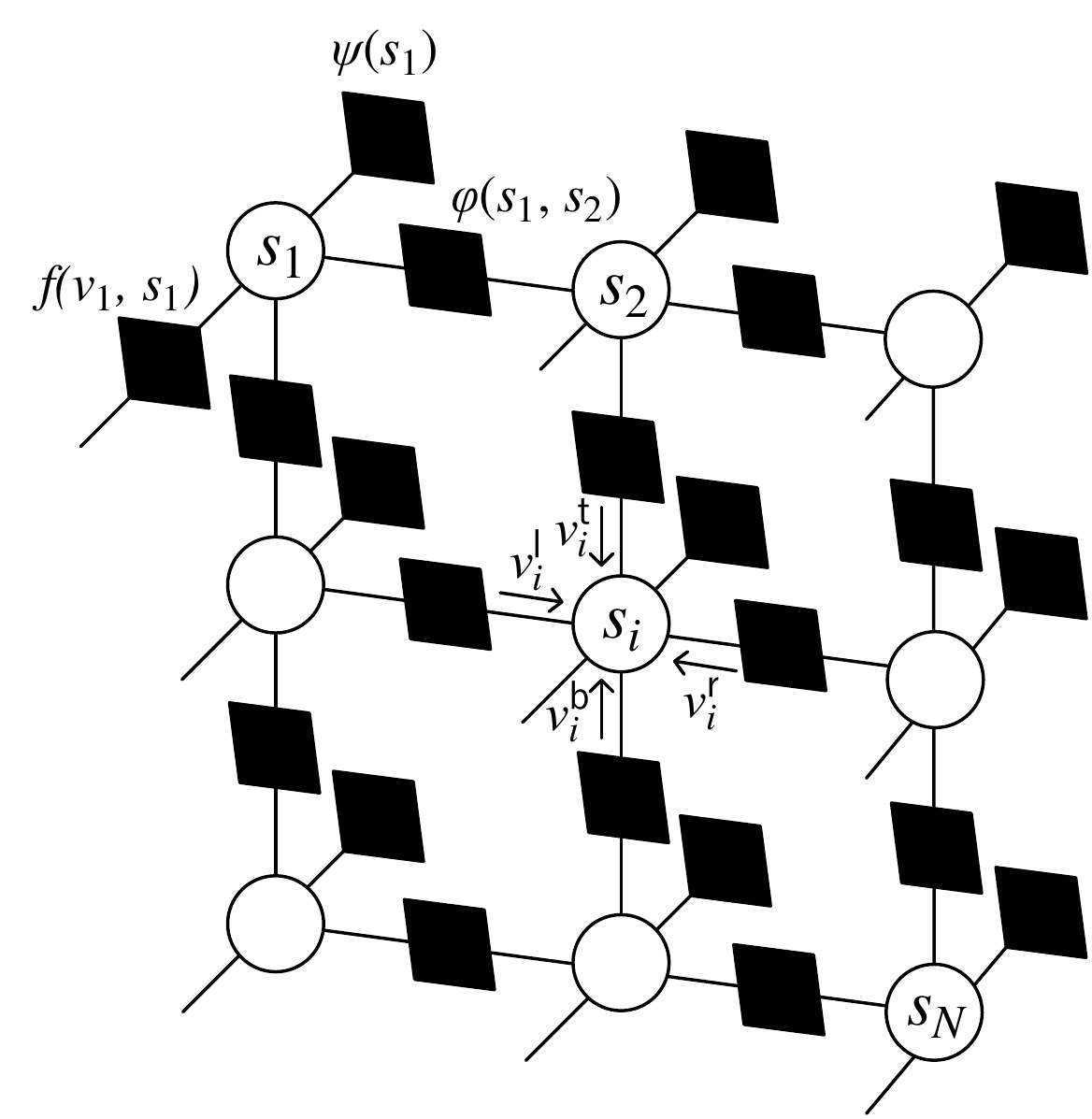}
    \caption{}
    \label{fg2b}
    \end{subfigure}
    \caption{(a) The factor graph characterizes the hierarchical probability model assumed in \eqref{Marginal}. (b) An example of the 4-connected MRF with dimension $3\times 3$.}
    \vspace{-0.2cm}
    \label{fg2}
\end{figure*}

\section{Probability Model}\label{ProbModel}
The goal of this work is to recover a block-sparse signal $\boldsymbol{x} \in \mathbb{C}^N$ from the noise-corrupted measurements
\begin{equation}\label{LinearModel}
    \boldsymbol{y} = \boldsymbol{Ax} + \boldsymbol{w}
\end{equation}
where $\boldsymbol{A} \in \mathbb{C}^{M \times N} ~ (M < N)$ is the measurement matrix, and $\boldsymbol{w} \in \mathbb{C}^{M}$ is a circularly symmetric complex Gaussian (CSCG) noise with zero mean and covariance matrix $\sigma^2 \boldsymbol{I}$. 
We use a hierarchical Gaussian prior model to characterize the block-sparse structure of the unknown signal $\bsm{x}$ in \eqref{LinearModel}. Specifically, $\boldsymbol{x}$ is assigned a conditional Gaussian prior
\begin{equation}
    p(\boldsymbol{x}|\boldsymbol{v}) = \prod_{i=1}^N p(x_i|v_i)
\end{equation}
where $\boldsymbol{x} = [x_1, \ldots, x_N]^T$, $\boldsymbol{v} = [v_1, \ldots, v_N]^T$, and $p(x_i|v_i) = \mathcal{CN}(x_i;0,v_i)$ is a CSCG distribution with zero mean and variance $v_i$. Note that each $v_i$ is the variance of signal component $x_i$ to control the sparsity. When $v_i$ approaches zero, the corresponding component $x_i$ becomes zero. In this work, each $v_i$ is assigned a conditionally independent distribution given by
\begin{equation}\label{pvisi}
    p(v_i|s_i) = \text{Gamma}(v_i;a,b)\delta(s_i-1)+\delta(v_i)\delta(s_i+1),
\end{equation}
where $s_i \in \{-1,1\}$ is a hidden binary state; $\delta (\cdot)$ denotes the Dirac delta function; $\text{Gamma}(v_i;a,b)$ is the Gamma distribution defined as
\begin{equation}\label{BG}
\text{Gamma}(v_i;a,b)=\left\{
\begin{aligned}
& \frac{b^a v_i^{a-1} e^{-bv_i}}{\Gamma(a)}, & ~ v_i>0 \\
& 0, & ~ \text{otherwise}
\end{aligned}
\right.
\end{equation}
with $\Gamma(a)=\int_0^\infty t^{a-1} e^{-t} \text{d} t$ being the Gamma function. We use the Gamma distribution to characterize the nonzero part of each nonnegative sparse random variable $v_i$. In addition, when there is no prior knowledge of the random variable, $a$ and $b$ can be set to a small value (e.g., $10^{-10}$) to make the distribution noninformative.
Let $\rho$ be the fraction of nonzero elements in $\boldsymbol{x}$, the distribution $p(s_i)$ is modeled as 
\begin{equation}
p(s_i)=\left\{
\begin{aligned}
& \rho,  ~~~~~~~~ s_i = 1  \\
& 1-\rho, ~ s_i = -1.
\end{aligned}
\right.
\end{equation}
Then, the marginal distribution of $v_i$ can be expressed as
\begin{align}
    p(v_i) & = \sum_{s_i} p(v_i|s_i)p(s_i) \notag \\
    & = \rho \text{Gamma}(v_i;a,b) + (1-\rho)\delta(v_i). \label{pvi}
\end{align}
Furthermore, we use a Markov random field (MRF) prior to describe the block-sparse structure of $\boldsymbol{v}$. The hidden state variables can be modeled by the classic Ising model \cite{som2011approximate} as
\begin{equation}\label{ps}
    p(\boldsymbol{s}) \propto \exp \left( \sum_{i=1}^N \left( \frac{1}{2}\sum_{k \in \mathcal{D}_i} \beta s_{k} - \alpha \right) s_i \right)
\end{equation}
where $\mathcal{D}_i \subset \{1,\ldots, N\} \backslash i$ is the neighbors of index $i$; $\alpha$ and $\beta$ are parameters of the MRF. A larger $\beta$ implies a larger size of each block of non-zeros, and a larger $\alpha$ encourages a sparser $\boldsymbol{x}$.

We proceed to perform Bayesian inference based on the proposed hierarchical model. From the Bayesian rule, the joint probability of $p(\boldsymbol{y}, \boldsymbol{x}, \boldsymbol{v}, \boldsymbol{s})$ can be decomposed as
\begin{align}\label{Marginal}
    p(\boldsymbol{y},\boldsymbol{x},\boldsymbol{v},\boldsymbol{s})
    & = p(\boldsymbol{y}|\boldsymbol{x}) p(\boldsymbol{x}|\boldsymbol{v}) p(\boldsymbol{v}|\boldsymbol{s}) p(\boldsymbol{s}) \notag \\
    & = p(\boldsymbol{y}|\boldsymbol{x})\prod_i^N p(x_i|v_i) p(v_i|s_i) p(\boldsymbol{s}).
\end{align}
The dependencies of the random variables in the factorization \eqref{Marginal} can be shown by a factor graph as depicted in Fig. \ref{fg2}(a), where circles represent variable nodes and squares represent factor nodes. The factor nodes $\{f_i\}$, $\{g_i\}$, $l$, and $\gamma$ in Fig. \ref{fg2}(a) are defined as
\begin{subequations}
\begin{align}
    f_i & : p(v_i|s_i), \\
    g_i & :  p(x_i|v_i)
     = \mathcal{CN}(x_i;0,v_i), \label{pxivi} \\
    l & : p(\boldsymbol{y}|\boldsymbol{x}) = \mathcal{CN}(\boldsymbol{y}-\boldsymbol{Ax};\boldsymbol{0},\sigma^2 \boldsymbol{I}), \\
    \gamma & : p(\boldsymbol{s}) \propto \left(\prod_{i=1}^N \prod_{k \in \mathcal{D}_i} \varphi(s_i,s_k) \right)^{\frac{1}{2}}\prod_{i=1}^N \psi(s_i) \label{gammaps}
\end{align}
\end{subequations}
where
\begin{subequations}
\begin{align}
      \varphi(s_i,s_k) & = \exp (\beta s_i s_k), \\
      \psi(s_i) & = \exp (-\alpha s_i).
\end{align}
\end{subequations}
The factor graph in Fig. 1(a) includes two modules, namely, the linear module that handles the linear constraint in \eqref{LinearModel} and the MRF module that handles the MRF prior of $\boldsymbol{s}$ in \eqref{ps}. We further see that, with $\gamma$ in \eqref{gammaps}, the MRF module can be expanded as a sub-factor graph. In this paper, we will mostly focus on the commonly used 4-connected MRF as illustrated in Fig. \ref{fg2b}, though our algorithm can be readily applied to other forms of MRFs including the one-dimensional Markov chain.

Based on the above probability model, an optimal solution of $\boldsymbol{x}$ can be found by solving $\max_{\boldsymbol{x}} p(\boldsymbol{x}|\boldsymbol{y})$. However, solving this problem is computationally infeasible even for moderate values of $M$ and $N$. In this paper, we propose a low-complexity yet near-optimal message passing algorithm, termed variance state propagation, as detailed in what follows.

\section{Variance State Propagation Algorithm}\label{SecVSP}

\subsection{Sum-Product Message Passing}\label{messagepassing}
We will basically follow the sum-product rule for message passing over the factor graph in Fig. \ref{fg2a}. We start from the output messages of the linear module. In Fig. \ref{fg2a}, suppose that variable node $v_i$ receives a message $\nu_{g_i \rightarrow v_i}$ from the factor node $g_i$. The message from $v_i$ to $f_i$ is still given by $\nu_{g_i \rightarrow v_i}$.  Then the message from $f_i$ to $s_i$ is a Bernoulli distribution given by
\begin{subequations}\label{nufisi}
    \begin{align}
        \nu_{f_i \rightarrow s_i} & \propto \int_{v_i} p(v_i|s_i) \nu_{g_i \rightarrow v_i} \\
        & = \pi_{f_i \rightarrow s_i} \delta(s_i-1) + (1-\pi_{f_i \rightarrow s_i})\delta(s_i+1)
    \end{align}
\end{subequations}
where
\begin{equation}\label{pifisi}
    \pi_{f_i \rightarrow s_i} = \frac{\int_{v_i} \nu_{g_i \rightarrow v_i} \text{Gamma}(v_i;a,b)}{\int_{v_i} \nu_{g_i \rightarrow v_i} \text{Gamma}(v_i;a,b) + \int_{v_i} \nu_{g_i \rightarrow v_i} \delta(v_i)}.
\end{equation}
With the inputs $\{\nu_{f_i \rightarrow s_i} \}$, we are now ready to describe the messages involved in the MRF. For simplicity, we give the details of the messages passed in the 4-connected MRF as shown in Fig. \ref{fg2b}. The left, right, top, bottom neighbors to node $s_i$ are indexed by $i_\mathsf{l}$, $i_\mathsf{r}$, $i_\mathsf{t}$, $i_\mathsf{b}$, respectively, i.e., $\mathcal{D}_i = \{i_\mathsf{l}, i_\mathsf{r}, i_\mathsf{t}, i_\mathsf{b}\}$. The left, right, top, and bottom input messages of each variable $s_{i}$ are represented as Bernoulli distributions $\nu_{i}^{\mathsf{l}},\ \nu_{i}^{\mathsf{r}},\ \nu_{i}^{\mathsf{t}}$, and $\nu_{i}^{\mathsf{b}}$, respectively. The input message of $s_{i}$ from the left is given by
\begin{align}\label{LeftMsg}
  \nu_{i}^{\mathsf{l}} & \propto \int_{s_{i_\mathsf{l}}}\nu_{f_{i_\mathsf{l}}\rightarrow s_{i_\mathsf{l}}}\prod_{k\in\{\mathsf{l},\mathsf{t},\mathsf{b}\}} \nu_{i_\mathsf{l}}^k  \psi(s_{i_\mathsf{l}}) \varphi(s_i,s_{i_\mathsf{l}}) \notag\\
  & = \lambda_{i}^{\mathsf{l}}\delta(s_{i}-1)+(1-\lambda_{i}^{\mathsf{l}})\delta(s_{i}+1)
\end{align}
where $\lambda_{i}^{\mathsf{l}}$ is shown in \eqref{lambdal} (at the top of the next page).
\begin{figure*}[t]
\begin{equation}\label{lambdal}
\lambda_{i}^{\mathsf{l}}=   \frac{\pi_{f_{i_\mathsf{l}} \rightarrow s_{i_\mathsf{l}}}\prod_{k \in\{\mathsf{l},\mathsf{t},\mathsf{b}\}}\lambda_{{i_\mathsf{l}}}^k e^{-\alpha+\beta}
  + (1-\pi_{f_{i_\mathsf{l}} \rightarrow s_{i_\mathsf{l}}})\prod_{k \in\{\mathsf{l},\mathsf{t},\mathsf{b}\}}(1-\lambda_{{i_\mathsf{l}}}^k)e^{\alpha-\beta}}
  {(e^\beta+e^{-\beta})\left(\pi_{f_{i_\mathsf{l}} \rightarrow s_{i_\mathsf{l}}}e^{-\alpha}\prod_{k \in\{\mathsf{l},\mathsf{t},\mathsf{b}\}}\lambda_{{i_\mathsf{l}}}^k
  +(1-\pi_{f_{i_\mathsf{l}} \rightarrow s_{i_\mathsf{l}}})e^\alpha\prod_{k \in\{\mathsf{l},\mathsf{t},\mathsf{b}\}}(1-\lambda_{{i_\mathsf{l}}}^k)\right)}
\end{equation}
\hrulefill
\end{figure*}
The messages from the right, the top, and the bottom have similar representations. The output message of the MRF for each $s_i$ can be calculated as
\begin{equation}\label{s2f}
  \nu_{s_{i}\rightarrow f_{i}}=\pi_{s_i \rightarrow f_i} \delta (s_{i}-1)+(1-\pi_{s_i \rightarrow f_i})\delta(s_{i}+1)
\end{equation}
where
\begin{equation}\label{piin}
  \pi_{s_i \rightarrow f_i}=\frac{e^{-\alpha} \prod_{k \in \{\mathsf{l},\mathsf{r},\mathsf{t},\mathsf{b}\}}\lambda_{i}^k}{e^{-\alpha} \prod_{k \in \{\mathsf{l},\mathsf{r},\mathsf{t},\mathsf{b}\}}\lambda_{i}^k + e^{\alpha} \prod_{k \in \{\mathsf{l},\mathsf{r},\mathsf{t},\mathsf{b}\}}(1-\lambda_{i}^k)}.
\end{equation}
Then, the message from $f_{i}$ to $v_{i}$ is a Bernoulli-Gamma distribution given by
\begin{subequations}
    \begin{align}
      \nu_{f_i \rightarrow v_i} &\propto \int_{s_i} p(v_i|s_i) \nu_{s_i \rightarrow f_i} \\
      & = \pi_{s_i \rightarrow f_i} \text{Gamma}(v_i;a,b) + (1-\pi_{s_i \rightarrow f_i})\delta(v_i). \label{vfivi}
    \end{align}
\end{subequations}
With $\nu_{v_i \rightarrow g_i} = \nu_{f_i \rightarrow v_i}$, the message from $g_i$ to $x_i$ is given by
\begin{equation}
    \nu_{g_i \rightarrow x_i} \propto \int_{v_i} p(x_i|v_i) \nu_{v_i \rightarrow g_i}.
\end{equation}
The message from $x_i$ to $l$ is $\nu_{x_i \rightarrow l}=\nu_{g_i \rightarrow x_i}$. We now consider the message from factor node $l$ back to variable node $x_i$. From the sum-product rule, $\nu_{l \rightarrow x_i}$ can be expressed as
\begin{align}
    \nu_{l \rightarrow x_i} & \propto \int_{\boldsymbol{x}_{\backslash i}} p(\boldsymbol{y}|\boldsymbol{x}) \prod_{i' \neq i} \nu_{x_{i'}\rightarrow l} \notag \\
    & = \int_{\boldsymbol{x}_{\backslash i}} p(\boldsymbol{y}|\boldsymbol{x}) \prod_{i' \neq i} \int_{v_{i'}} p(x_{i'}|v_{i'}) \nu_{v_{i'} \rightarrow g_{i'}} \label{integration}
\end{align}
for $i' \neq i$, where $\boldsymbol{x}_{\backslash i}$ denotes all the entries of $\boldsymbol{x}$ except the $i$-th entry. Clearly, $\nu_{x_i \rightarrow g_i} = \nu_{l \rightarrow x_i}, \forall i$. Then, the messages $\nu_{g_i \rightarrow v_i}$ and $\nu_{v_i \rightarrow f_i}$ can be computed as
\begin{equation}\label{givi}
    \nu_{v_i \rightarrow f_i} = \nu_{g_i \rightarrow v_i} \propto \int_{x_i} \nu_{x_i \rightarrow g_i} p(x_i|v_i)
\end{equation}
where $\nu_{x_i \rightarrow g_i} = \nu_{l \rightarrow x_i}$. The above messages are calculated iteratively until convergence.
\subsection{Update of $\nu_{g_i \rightarrow v_i}$}\label{updatevi}
The algorithm in Subsection \ref{messagepassing} is a straightforward application of the sum-product rule for  message passing. This algorithm, however, is difficult to implement due to the high computational complexity involved in evaluating the integrals in \eqref{integration} and \eqref{givi}. To reduce complexity, we propose to approximate the outputs of the linear module $\{ \nu_{g_i \rightarrow v_i} \}$ as follows. By substituting $\nu_{x_i \rightarrow g_i} = \nu_{l \rightarrow x_i}$ in \eqref{integration}, the message $\nu_{g_i \rightarrow v_i}$ in \eqref{givi} can be written as
\begin{align}
    \nu_{g_i \rightarrow v_i} & \propto \int_{x_i} p(x_i|v_i)  \int_{\boldsymbol{x}_{\backslash i}} p(\boldsymbol{y}|\boldsymbol{x})  \int_{\boldsymbol{v}_{\backslash i}} \prod_{i' \neq i } \left( \nu_{v_{i'} \rightarrow g_{i'}}p(x_{i'}|v_{i'}) \right)  \notag \\
    & \propto \int_{\boldsymbol{v}_{\backslash i}} \prod_{i' \neq i } \nu_{v_{i'} \rightarrow g_{i'}} \left( \int_{\boldsymbol{x}} p(\boldsymbol{y}|\boldsymbol{x}) p(\boldsymbol{x}|\boldsymbol{v}) \right) \notag \\
    & \propto \int_{\boldsymbol{v}_{\backslash i}} p(\boldsymbol{y}|\boldsymbol{v}) \prod_{i' \neq i } \nu_{v_{i'} \rightarrow g_{i'}}.  \label{viinte}
\end{align}
The integral in \eqref{viinte} is difficult to solve.
To simplify the message calculation, we propose to replace the output of the linear module for node $v_i$ by the mean $\mu_{g_i \rightarrow v_i} = \mathbb{E}_{\nu_{g_i \rightarrow v_i}}[v_i]$, where the expectation $\mathbb{E}$ is taken over the distribution $\nu_{g_i \rightarrow v_i}$. Similar ideas for message replacements and approximations have been previously employed, e.g., in denoising-based turbo compressed sensing \cite{xue2017denoising} in which a denoiser is used for message approximation when the probability model of a node is incomplete or unavailable.
While $\mathbb{E}_{\nu_{g_i \rightarrow v_i}}[v_i]$ is still difficult to evaluate, we further approximate $\mu_{g_i \rightarrow v_i}$ by
\begin{equation}\label{mugivit1}
    \mu_{g_i \rightarrow v_i} = \argmax_{v_i} p(\boldsymbol{y}|\boldsymbol{v})|_{\{ v_{i'}=\mu_{v_{i'} \rightarrow g_{i'}}, \forall i' \neq i \}},
\end{equation}
where $\mu_{v_i \rightarrow g_i}$ is the input mean of $v_i$ for the linear module, i.e.
\begin{equation}\label{muvigi}
    \mu_{v_i \rightarrow g_i} = \mathbb{E}_{\nu_{v_i \rightarrow g_i}}[v_i],  \forall i.
\end{equation}
It is interesting to compare \eqref{mugivit1} with \eqref{viinte}. We may treat each $ \nu_{v_{i'} \rightarrow g_{i'}}$ as the prior of $v_{i'}$, for $i' \neq i$. Then $\nu_{g_i \rightarrow v_i}$ in \eqref{viinte} can be regarded as the likelihood of $v_i$ given $\boldsymbol{y}$, and $\mathbb{E}_{\nu_{g_i \rightarrow v_i}}[v_i]$ is the corresponding mean. In contrast, \eqref{mugivit1} only requires the prior mean $\mu_{v_{i'} \rightarrow g_{i'}}$ of each $v_{i'}$ rather than the whole distribution $\nu_{v_{i'} \rightarrow g_{i'}}$. Thus, $\mu_{g_i \rightarrow v_i}$ in \eqref{mugivit1} can be treated as the maximum likelihood of $v_i$ given $\boldsymbol{y}$ and $\mu_{v_{i'} \rightarrow g_{i'}},~ i'\neq i$. From the estimation theory, it is known that $\mu_{g_i \rightarrow v_i}$ given by \eqref{mugivit1} is always inferior to $\mu_{g_i \rightarrow v_i}$ calculated based on \eqref{viinte}, provided that the prior distributions $\{\nu_{v_i' \rightarrow g_i'} \}$ are accurate. We next present two methods to solve $\mu_{g_i \rightarrow v_i}$ in \eqref{mugivit1}. For notational convenience, denote
\begin{subequations}
\begin{align}
    \mathcal{S} & = \{ v_{i}=\mu_{v_{i} \rightarrow g_{i}}, \forall i \} \\
    \mathcal{S}_{\backslash i} & = \{ v_{i'}=\mu_{v_{i'} \rightarrow g_{i'}}, \forall i' \neq i \}.
\end{align}
\end{subequations}

\subsection{Gradient Method for Solving \eqref{mugivit1}}\label{SecGDS}
For the problem in \eqref{mugivit1}, a straightforward  solution is to find a stationary point of $p(\boldsymbol{y}|\boldsymbol{v})|_{\mathcal{S}_{\backslash i}}$ with respect to $v_i$ via gradient descent (GD). We note that $p(\boldsymbol{x}|\boldsymbol{y},\boldsymbol{v}) \propto p(\boldsymbol{y}|\boldsymbol{x}) p(\boldsymbol{x}|\boldsymbol{v})$ is a complex Gaussian distribution with the mean $\boldsymbol{m}$ and the covariance  $\boldsymbol{\Phi}$ given by
\begin{align}
    \boldsymbol{m} & = \sigma^{-2} \boldsymbol{\Phi} \boldsymbol{A}^{{H}} \boldsymbol{y} \label{m}\\
    \boldsymbol{\Phi} & = \left(\sigma^{-2} \boldsymbol{A}^{{H}} \boldsymbol{A}+\boldsymbol{D}^{-1}\right)^{-1} \label{phi}
\end{align}
where $\boldsymbol{D}$ is a diagonal matrix with the $i$-th diagonal element equal to $v_i$. Then, $p(\boldsymbol{y}|\boldsymbol{v})$ can be expressed as
\begin{align}
    p(\boldsymbol{y}|\boldsymbol{v}) & = \int_{\boldsymbol{x}} p(\boldsymbol{y}|\boldsymbol{x}) p(\boldsymbol{x}|\boldsymbol{v}) \notag \\
    & = \frac{\exp \left( \boldsymbol{m}^{H} \boldsymbol{\Phi}^{-1} \boldsymbol{m} - \sigma^{-2}\boldsymbol{y}^{H}\boldsymbol{y}\right) |\boldsymbol{\Phi}|}{(\pi \sigma^2)^M |{\boldsymbol{D}}|} \int_{\boldsymbol{x}} \mathcal{CN}(\boldsymbol{x};\boldsymbol{m},\boldsymbol{\Phi}) \notag \\
    & = \frac{\exp \left( -\sigma^{-2} \boldsymbol{y}^{H} \boldsymbol{y} \right)}{(\pi \sigma^2)^M} \cdot \frac{ \exp \left(\boldsymbol{m}^{H} \boldsymbol{\Phi}^{-1} \boldsymbol{m} \right) |\boldsymbol{\Phi}|}{\prod_{i=1}^N v_i}.  \label{pyvintegration}
\end{align}
The first term in \eqref{pyvintegration} is independent of $\boldsymbol{v}$.
Thus, problem \eqref{mugivit1} can be equivalently written as
\begin{equation}\label{argchi}
    \mu_{g_i \rightarrow v_i} = \argmin_{v_i} \chi (\boldsymbol{y}, \boldsymbol{v})|_{\mathcal{S}_{\backslash i}},
\end{equation}
where
\begin{align}
    \chi (\boldsymbol{y}, \boldsymbol{v}) & \triangleq - \ln \frac{ \exp \left(\boldsymbol{m}^{H} \boldsymbol{\Phi}^{-1} \boldsymbol{m} \right) |\boldsymbol{\Phi}|}{\prod_{i=1}^N v_i} \notag \\
    & = - \boldsymbol{m}^{H} \boldsymbol{\Phi}^{-1} \boldsymbol{m} - \ln |\boldsymbol{\Phi}| + \sum_{i=1}^N \ln v_i . \label{chi}
\end{align}
It is difficult to obtain an analytical solution to problem \eqref{argchi}. We propose to use the gradient descent to find a stationary point of \eqref{argchi}, with the update rule given by
\begin{equation}\label{updaterule}
    \mu_{g_i \rightarrow v_i} = \mu_{v_i \rightarrow g_i} - \epsilon_i \frac{\partial \chi}{\partial v_i}|_{v_i = \mu_{v_i \rightarrow g_i}}
\end{equation}
where $\epsilon_i$ is an appropriate step size that can be selected from the backtracking line search to satisfy
\begin{equation}
    \chi (\boldsymbol{y}, \boldsymbol{v})|_{\mathcal{S}_{\backslash i}, v_i = \mu_{g_i \rightarrow v_i}} \leq \chi (\boldsymbol{y}, \boldsymbol{v})|_{\mathcal{S}}.
\end{equation}
The closed-form expression of the partial derivative $\frac{\partial \chi}{\partial v_i}$ is given by
\begin{equation}\label{pchi}
    \frac{\partial \chi}{\partial v_i} = - \left( \frac{\boldsymbol{y}^{H} \boldsymbol{A} \boldsymbol{u}_i}{\sigma^{2} v_i}\right)^2 - \text{Tr} \left[ \boldsymbol{E}_i \boldsymbol{\Phi} \right] + \frac{1}{v_i},
\end{equation}
where $\boldsymbol{u}_i$ is the $i$-th column of $\boldsymbol{\Phi}$, and
\begin{equation}\label{Ei}
    \boldsymbol{E}_i \triangleq
    \begin{bmatrix}
        \boldsymbol{0} \\
        & \frac{1}{v_i^2} \\
        & & \boldsymbol{0}
    \end{bmatrix}
\end{equation}
is a diagonal matrix with only one nonzero element $\frac{1}{v_i^2}$ in the $i$-th diagonal position. The detailed derivation of \eqref{pchi} is given in Appendix \ref{apdixa}. We can update all the entries of $\boldsymbol{v}$ based on \eqref{updaterule} in a sequential manner. However, when $N$ is large, using backtracking to calculate $\epsilon_i$ for every single $v_i$ sequentially imposes a heavy computational burden. We consider to use a common step size for all $\{ v_i \}$, i.e.
\begin{equation}\label{gradmunew}
    \boldsymbol{\mu}^{(\text{new})} = \boldsymbol{\mu}^{(\text{old})} - \epsilon \nabla \chi (\boldsymbol{y}, \boldsymbol{\mu}^{(\text{old})})
\end{equation}
where $\boldsymbol{\mu}^{(\text{old})} \triangleq \left[\mu_{v_1 \rightarrow g_1}, \ldots, \mu_{v_N \rightarrow g_N} \right]^T$,
the gradient $\nabla \chi(\boldsymbol{y}, \boldsymbol{v})$ is defined by
\begin{equation}\label{nablachi}
    \nabla \chi(\boldsymbol{y}, \boldsymbol{v}) \triangleq \left[ \frac{\partial \chi}{\partial v_1}, \ldots, \frac{\partial \chi}{\partial v_N} \right]^T,
\end{equation}
and $\epsilon$ is the common step size obtained from the backtracking line search rule satisfying
\begin{equation}\label{LineSearchCond}
    \chi (\boldsymbol{y}, \boldsymbol{\mu}^{(\text{new})}) \leq \chi (\boldsymbol{y}, \boldsymbol{\mu}^{(\text{old})}).
\end{equation}
Thus, by letting $[\mu_{g_1 \rightarrow v_1}, \ldots, \mu_{g_N \rightarrow v_N}] = \boldsymbol{\mu}^{(\text{new})}$, we obtain an update of $\{\mu_{g_i \rightarrow v_i}\}$. The above approximate solution to \eqref{viinte} is referred to as the GD-based solver, as summarized in Algorithm \ref{GDS}. We note that iteration is introduced in Algorithm \ref{GDS} to find a stationary point of \eqref{mugivit1}.

\subsection{ELBO-Based Method for Solving \eqref{mugivit1}}\label{SecEMS}
The gradient descent method described above, though conceptually simple, suffers from slow convergence and high complexity. This inspires us to develop an alternative solution to \eqref{mugivit1} with improved performance. Let $q(\boldsymbol{x})$ be a distribution function of $\boldsymbol{x}$, and define
\begin{align}
    \eta(v_i) & \triangleq \ln p(\boldsymbol{y}|\boldsymbol{v})|_{\mathcal{S}_{\backslash i}  } \notag \\
    & = \mathcal{L}(v_i,q(\boldsymbol{x})) + D_{\text{KL}}(q(\boldsymbol{x})||p(\boldsymbol{x}|\boldsymbol{y},\boldsymbol{v})|_{\mathcal{S}_{\backslash i}  }) \label{etavi}
\end{align}
where $\mathcal{L}(v_i,q(\boldsymbol{x}))$ is the evidence lower bound (ELBO) defined by
\begin{equation}
    \mathcal{L}(v_i,q(\boldsymbol{x})) \triangleq \int_{\boldsymbol{x}}q(\boldsymbol{x})\ln \frac{p(\boldsymbol{y},\boldsymbol{x}|\boldsymbol{v})|_{\mathcal{S}_{\backslash i}  }}{q(\boldsymbol{x})},
\end{equation}
and
\begin{multline}
    D_{\text{KL}}(q(\boldsymbol{x})||p(\boldsymbol{x}|\boldsymbol{y},\boldsymbol{v})|_{\mathcal{S}_{\backslash i}  }) \\ \triangleq - \int_{\boldsymbol{x}}q(\boldsymbol{x})\ln \frac{p(\boldsymbol{x}|\boldsymbol{y},\boldsymbol{v})|_{\mathcal{S}_{\backslash i}  }}{q(\boldsymbol{x})}
\end{multline}
is the Kullback-Leibler divergence between $q(\boldsymbol{x})$ and $p(\boldsymbol{x}|\boldsymbol{y},\boldsymbol{v})|_{\mathcal{S}_{\backslash i}}$. Equation \eqref{etavi} holds for any choice of $q(\boldsymbol{x})$ and $v_i$. Since $D_{\text{KL}}(q(\boldsymbol{x})||p(\boldsymbol{x}|\boldsymbol{y},\boldsymbol{v})|_{\mathcal{S}_{\backslash i}}) \geq 0$, $\mathcal{L}(v_i,q(\boldsymbol{x}))$ is indeed a lower bound of $\eta(v_i)$. Thus,
to approximately solve \eqref{mugivit1}, we turn to maximize $\mathcal{L}(v_i,q(\boldsymbol{x}))$ as
\begin{equation}\label{L}
    \mu_{g_i \rightarrow v_i} = \argmax_{v_i} \mathcal{L}(v_i,q(\boldsymbol{x}))
\end{equation}
where $q(\boldsymbol{x})$ is chosen as
\begin{equation}\label{tildeqx}
    q  (\boldsymbol{x}) = p(\boldsymbol{x}|\boldsymbol{y},\boldsymbol{v})|_{\mathcal{S}  }.
\end{equation}
The following proposition ensures that the choice of $q(\boldsymbol{x})$ in \eqref{tildeqx} yields a good approximate solution to \eqref{mugivit1}.
\begin{prop}\label{prop1}
With $\mu_{g_i \rightarrow v_i}$ given by \eqref{L}, the following inequality holds:
\begin{equation}\label{prop1eq}
    p(\boldsymbol{y}|\boldsymbol{v})|_{ \mathcal{S}_{\backslash i}, v_i = \mu_{g_i \rightarrow v_i} } \geq p(\boldsymbol{y}|\boldsymbol{v})|_{\mathcal{S}}.
\end{equation}
\end{prop}
The proof of Proposition \ref{prop1} is given in Appendix \ref{apdixb}.
The following Proposition gives the solution of \eqref{L}.
\begin{prop}\label{prop2}
The solution of \eqref{L} is given by
\begin{equation}\label{alphaupdate}
    \mu_{g_i \rightarrow v_i} = |m_i|^2 + \phi_{i,i},
\end{equation}
where $m_i$ denotes the $i$-th entry of $\boldsymbol{m}$ in \eqref{m}, and $\phi_{i,i}$ denotes the $i$-th diagonal element of the covariance matrix $\boldsymbol{\Phi}$ in \eqref{phi}.
\end{prop}
The proof of Proposition \ref{prop2} is given in Appendix \ref{apdixc}. In this way, we obtain an update of $\mu_{g_i \rightarrow v_i}$ for each $i$. The above approximate solution to \eqref{viinte} is referred to as the ELBO-based solver as summarized in Algorithm \ref{EMS}. Similarly to the GD-based solver, iteration is introduced to ensure a better performance.

\subsection{Update of $\pi_{f_i \rightarrow s_i}$}\label{pifisisec}
Recall from \eqref{mugivit1} that we replace the message $\nu_{g_i \rightarrow v_i}$ by the mean $\mu_{g_i \rightarrow v_i}$. As a consequence, $\pi_{f_i \rightarrow s_i}$ in \eqref{nufisi} cannot be calculated by using \eqref{pifisi} since $\nu_{g_i \rightarrow v_i}$ is not available. To carry out message passing from $f_i$ to $s_i$, we need to find a new approach to update $\pi_{f_i \rightarrow s_i}$.

In \eqref{muvigi} we notice that $\mu_{v_i \rightarrow g_i}$ is set to the mean of $v_i \sim \nu_{v_i \rightarrow g_i}$, and $\nu_{v_i \rightarrow g_i}$ is a Bernoulli-Gamma distribution shown in \eqref{vfivi}.
Thus, $\mu_{v_{i} \rightarrow g_{i}}$ is calculated by using
\begin{equation}\label{mapfivi}
    \mu_{v_{i} \rightarrow g_{i}} = \mathbb{E}_{\nu_{v_{i} \rightarrow g_{i}}}[v_{i}] = \frac{a}{b} \pi_{s_i \rightarrow f_i},
\end{equation}
where $\frac{a}{b}$ is the mean of the Gamma distribution $\text{Gamma}(v_i;a,b)$. Equation \eqref{mapfivi} shows that under the probability model specified in Section \ref{ProbModel}, $\mu_{v_{i} \rightarrow g_{i}}$ is simply the product of $\pi_{s_i \rightarrow f_i}$ and the mean $\frac{a}{b}$ of $\text{Gamma}(v_i;a,b)$. Inspired by this, we propose a \textit{moment matching} method that mimics the relationship between $\pi_{s_i \rightarrow f_i}$ and $\mu_{v_i \rightarrow g_i}$ to establish a map between $\mu_{g_i \rightarrow v_i}$ and $\pi_{f_i \rightarrow s_i}$, i.e.
\begin{equation}\label{barpi}
    \mu_{g_i \rightarrow v_i} = \frac{a}{b}\pi_{f_i \rightarrow s_i}.
\end{equation}
Here, for $\mu_{g_i \rightarrow v_i}$ given by the linear module, $\pi_{s_i \rightarrow f_i}$ may be greater than $1$ and therefore is not necessarily a valid probability. To avoid this, we need to limit $\pi_{s_i \rightarrow f_i}$ to the range of $[0,1]$.
That is, given $\mu_{g_i \rightarrow v_i}$ and the parameters $\{a,b\}$ of $\text{Gamma}(\cdot;a,b)$, $\pi_{f_i \rightarrow s_i}$ can be approximated as
\begin{align}\label{xiinv}
    \pi_{f_i \rightarrow s_i}
    & = \min \left(\frac{\mu_{g_i \rightarrow v_i}}{a/b},1 \right),
\end{align}
where $\pi_{f_i \rightarrow s_i}$ is cropped to 1 to ensure that it is a valid probability. Then $\nu_{f_i \rightarrow s_i}$ in \eqref{nufisi} can be constructed by using $\pi_{f_i \rightarrow s_i}$ in \eqref{xiinv}. Subsequent message passing can therefore proceed.


\subsection{Parameter Tuning}\label{ParaTuning}
In this subsection, we discuss the choice of model parameters. The noise variance $\sigma^2$ is considered as a \textit{priori} known in the previous sections. In practice, $\sigma^2$ can be learned under the expectation-maximization (EM) framework in a similar way as in the SBL \cite{tropp2007signal}.

\begin{algorithm}[t]
\LinesNumbered

\KwIn{$\boldsymbol{y}$, $\boldsymbol{A}$, $\{\mu_{v_i \rightarrow g_i}\}$, $T_{\normalfont \rm{in}}$, $\sigma^2$}
$t = 0$\;
$\hat{\boldsymbol{\mu}} = [\mu_{v_1 \rightarrow g_1}, \ldots, \mu_{v_N \rightarrow g_N}]^T$\;
\While{t $< T_{\normalfont \rm{in}}$}{
Compute $\nabla \chi (\boldsymbol{y}, \hat{\boldsymbol{\mu}})$ based on \eqref{pchi}\;
Find $\epsilon$ that satisfies $\chi \left(\boldsymbol{y},\hat{\boldsymbol{\mu}} + \epsilon \nabla \chi(\boldsymbol{y}, \hat{\boldsymbol{\mu}}) \right) \leq \chi(\boldsymbol{y},\hat{\boldsymbol{\mu}})$ using the backtracking line search\;
$\hat{\boldsymbol{\mu}} = \hat{\boldsymbol{\mu}} - \epsilon \nabla \chi(\boldsymbol{y}, \hat{\boldsymbol{\mu}})$\;
$t = t+1$\;
}
\KwOut{$[\mu_{g_1 \rightarrow v_1}, \ldots, \mu_{g_N \rightarrow v_N}]^T = \hat{\boldsymbol{\mu}}$}
\caption{GD-based solver for \eqref{mugivit1}}
\label{GDS}
\end{algorithm}

\begin{algorithm}[t]
\LinesNumbered


\KwIn{$\boldsymbol{y}$, $\boldsymbol{A}$, $\{\mu_{v_i \rightarrow g_i}\}$, $T_{\normalfont \rm{in}}$, $\sigma^2$}

$t = 0$\;
$\hat{\boldsymbol{\mu}}^{(0)} = [\mu_{v_1 \rightarrow g_1}, \ldots, \mu_{v_N \rightarrow g_N}]^T$\;
\While{t $< T_{\normalfont \rm{in}}$}{
$\boldsymbol{D} = \rm{diag}(\hat{\boldsymbol{\mu}})$\;
Compute $\boldsymbol{m}$ and $\boldsymbol{\Phi}$ according to \eqref{m} and \eqref{phi}\;
Update $\hat{\boldsymbol{\mu}} = \left[\hat{\mu}_{g_1 \rightarrow v_1}, \ldots, \hat{\mu}_{g_N \rightarrow v_N} \right]^T$ with each $\hat{\mu}_{g_i \rightarrow v_i} = |m_i|^2 + \phi_{i,i}$\;
$t = t+1$\;
}
\KwOut{$[\mu_{g_1 \rightarrow v_1}, \ldots, \mu_{g_N \rightarrow v_N}]^T = \hat{\boldsymbol{\mu}}$}
\caption{ELBO-based solver for \eqref{mugivit1}}
\label{EMS}
\end{algorithm}

Parameters $\alpha$ and $\beta$ of the MRF can also be learned by the EM algorithm. However, we find in numerical experiments that  the algorithm performs well when $\alpha$ and $\beta$ are fixed to empirical values, and learning $\alpha$ and $\beta$ by EM yields a marginal gain.

\begin{algorithm}[t]\label{VSPAlgo}
\LinesNumbered

\KwIn{$\boldsymbol{y}$, $\boldsymbol{A}$, $a$, $b$, $T_{\rm{out}}$, $T_{\normalfont \rm{in}}$, $\alpha$, $\beta$, $\rho$, $\vartheta$, $\sigma^2$}
$t = 0$, $\mu_{v_i \rightarrow g_i} = 0, \forall i$\;
\While{t $< T_{\rm{out}}$}{
Call Algorithm \ref{GDS} or \ref{EMS} to update $\{\mu_{g_i \rightarrow v_i}\}$\;
\If{$t<T-1$}
{
    Compute $\kappa$ based on \eqref{kappa}\;
    $\forall i \in \{1,\ldots,N\}$, compute $\pi_{f_i \rightarrow s_i}$ based on \eqref{pinew}\;
    $\forall i \in \{1,\ldots,N\}$, compute $\pi_{s_i \rightarrow f_i}$ based on \eqref{piin}\;
    $\forall i \in \{1,\ldots,N\}$, compute $\mu_{v_i \rightarrow g_i}$ based on \eqref{munew}\;
}
$t = t+1$\;
}
Compute the posterior mean $\boldsymbol{m}$ of $\boldsymbol{x}$ in \eqref{m} based on $\{\mu_{g_i \rightarrow v_i}\}$\;
\KwOut{$\hat{\boldsymbol{x}} = \boldsymbol{m}$}
\caption{VSP algorithm}
\label{Algo1}
\end{algorithm}

The update of the two parameters $a$ and $b$ of the check function $f_i$ in \eqref{pvisi} is more crucial. At the beginning of the iteration, since there is not much prior information of the sparse signal $\boldsymbol{x}$, we set $a$ and $b$ to very small values (e.g., $a=b=10^{-10}$) to ensure that the Gamma distribution part in $f_i$ is noninformative. As the iteration proceeds, we update $a$ and $b$ to make the calculation of the messages more accurate. In \eqref{mapfivi} and \eqref{xiinv}, we notice that when calculating $\mu_{v_{i} \rightarrow g_{i}}$ and $\pi_{f_i \rightarrow s_i}$, only $a/b$ (the mean of $\text{Gamma}(v_i;a,b)$) is used. So in the update of $f_i$, we fix $a$ and only update $b$ to adjust the mean of $\text{Gamma}(v_i;a,b)$.
In iteration, each $v_i$ receives an update $\mu_{g_i \rightarrow v_i}$ that is an estimate of $v_i$. From the probability model, we see that $\{v_i\}$ are drawn from \eqref{pvi}. Thus, with high probability, $v_i$s with the largest values of $\mu_{g_i \rightarrow v_i}$ are nonzeros drawn from the Gamma distribution part $\text{Gamma}(v_i;a,b)$ in $f_i$. Based on this observation, we propose to update $\frac{a}{b}$ as follows. Let $\mu_1 \geq \mu_2 \geq \ldots \geq \mu_N$ be the reordered sequence of $\mu_{g_1 \rightarrow v_1}, \ldots, \mu_{g_N \rightarrow v_N}$. We assume that those $v_i$s corresponding to the ${K'}$ largest entries, i.e., $\{\mu_1, \ldots, \mu_{K'} \}$, are drawn from $\text{Gamma}(v_i;a,b)$. Then, by approximating the statistical mean by a sample mean, we obtain
\begin{equation}\label{kappa}
    \frac{a}{b} = \kappa \triangleq \frac{1}{{K'}} \sum_{h=1}^{K'} \mu_h.
\end{equation}
For the choice of ${K'}$, it is found in the experiments that the best performance is achieved when ${K'}= \lfloor \vartheta K \rceil$ with coefficient $\vartheta \in [1,2]$, where $\lfloor u \rceil$ is a function which returns the nearest integer to $u \in \mathbb{R}$, and $K$ is the number of nonzero elements in $\boldsymbol{x}$. Based on the above discussion, the calculations of $\pi_{f_i \rightarrow s_i}$ and $\mu_{v_{i} \rightarrow g_{i}}$ are changed to
\begin{align}
    \pi_{f_i \rightarrow s_i} & = \min \left(\frac{\mu_{g_i \rightarrow v_i}}{\kappa},1 \right) \label{pinew} \\
    \mu_{v_{i} \rightarrow g_{i}} & = \kappa \pi_{s_i \rightarrow f_i}. \label{munew}
\end{align}

\subsection{Overall Algorithm}

\renewcommand\arraystretch{1.25}
\begin{table*}[t]
    \centering
    \caption{Properties of VSP and Other Popular Compressed Sensing Algorithms}
        \begin{threeparttable}
        \begin{tabular}{l c p{1.3cm}<{\centering} p{1.2cm}<{\centering} p{1.3cm}<{\centering} c c l}
       \toprule
       \textbf{Algorithm} & \textbf{Complexity}\footnotemark[1] & \textbf{Number of Blocks}\footnotemark[2] & \textbf{Size of Blocks}\footnotemark[2] & \textbf{Location of Blocks}\footnotemark[2] & \textbf{Sparsity}\footnotemark[2] & \textbf{Robustness}\footnotemark[3] & \textbf{Description} \\
        \midrule
        OMP \cite{tropp2007signal} & $\mathcal{O}(KMN)$ & -- & -- & -- & \cmark & Yes & \multirow{2}{5cm}{Naive greedy algorithms that require a large number of measurements for reliable recovery.} \\[1ex]
        CoSaMP \cite{needell2009cosamp} & $\mathcal{O}(TMN)$ & -- & -- & -- & \cmark & Yes & \\
        \midrule
        SBL \cite{tipping2001sparse} & $\mathcal{O}(T M^3)$ & -- & -- & -- & \xmark & Yes & \multicolumn{1}{m{5cm}}{A Bayesian-based algorithm that repeatedly calculates the posterior distribution of the signal and updates the parameters of the model prior.} \\
        \midrule
        IHT \cite{blumensath2009iterative} & $\mathcal{O}(TMN)$ & -- & -- & -- & \cmark & Yes & \multicolumn{1}{m{5cm}}{A representative iterative thresholding algorithm.} \\
        \midrule
        Block-OMP \cite{eldar2010block} & -- & \cmark & \cmark & \cmark & \cmark & Yes & \multirow{3}{5cm}{Improved greedy algorithms specified for block-sparse signal recovery. It requires strong prior knowledge.} \\
        Block-CoSaMP \cite{eldar2010block} & -- & \cmark & \cmark & \cmark & \cmark & Yes & \\
        Struct-OMP \cite{huang2011learning} & -- & \cmark & \xmark & \xmark & \cmark & Yes & \\
        \midrule
        CluSS--MCMC \cite{yu2012bayesian} & $\mathcal{O}(T N^2)$ & \xmark & \xmark & \xmark & \xmark & Yes & \multicolumn{1}{m{5cm}}{A Bayesian based algorithm that adopts a Bernoulli-Gaussian hierarchical model as the prior. Since the posterior distribution can not be derived analytically, MCMC sampling is employed in inference.} \\
        \midrule
        PC-SBL \cite{fang2014pattern} & $\mathcal{O}(T M^3)$ & \xmark & \xmark & \xmark & \xmark & Yes & \multicolumn{1}{m{5cm}}{An improved algorithm specified for block-sparse signals based on the SBL framework. A new pattern-coupled Gaussian probability model is employed as the prior.} \\
        PCSBL-GAMP \cite{fang2016two} & $\mathcal{O}(TMN)$ & \xmark & \xmark & \xmark & \xmark & No & \multicolumn{1}{m{5cm}}{A modified PC-SBL algorithm which reduces the complexity of PC-SBL by using message passing, but at the same time increases the sensitivity to the measurement matrix.} \\       
        \midrule        
        STCS \cite{chen2017structured} & $\mathcal{O}(T M^3)$ & \xmark & \xmark & \xmark & \xmark & No & \multirow{2}{5cm}{Message passing algorithms that have superior performance when the measurement matrix satisfies certain conditions, but have no performance guarantee when the conditions are not met.}
        \\[6ex]
        CGAMP \cite{he2019super} & $\mathcal{O}(TMN)$ & \xmark & \xmark & \xmark & \xmark & No & \\
        \midrule
        VSP & $\mathcal{O}(TM^3)$ & \xmark & \xmark & \xmark & \xmark & Yes & \multicolumn{1}{m{5cm}}{A new message-passing based algorithm that has near-optimal performance and is robust to the choice of the measurement matrix.} \\
        \bottomrule
        \end{tabular}
        \begin{tablenotes}
        \item[1] ``Complexity'' here is evaluated for a general measurement matrix. The complexities of some algorithms (such as STCS) can be reduced when the measurement matrix takes a special structure. ``$T$'' stands for the total number of iterations. Particularly, $T=T_{\text{in}} T_{\text{out}}$ for VSP. 
        \item[2] ``\cmark'' denotes ``necessary'' for the corresponding algorithm. ``\xmark'' denotes ``unnecessary'' for the corresponding algorithm. ``--'' denotes no consideration for the corresponding algorithm. 
        \item[3] ``Robustness'' here means the robustness of the algorithm to the choice of the measurement matrix.
        \end{tablenotes}
        \end{threeparttable}
        \vspace{-0.2cm}
    \label{table1}
\end{table*}

\begin{figure*}[t]
    \centering
    \begin{subfigure}{0.48\linewidth}
    \centering
    \includegraphics[width=\linewidth]{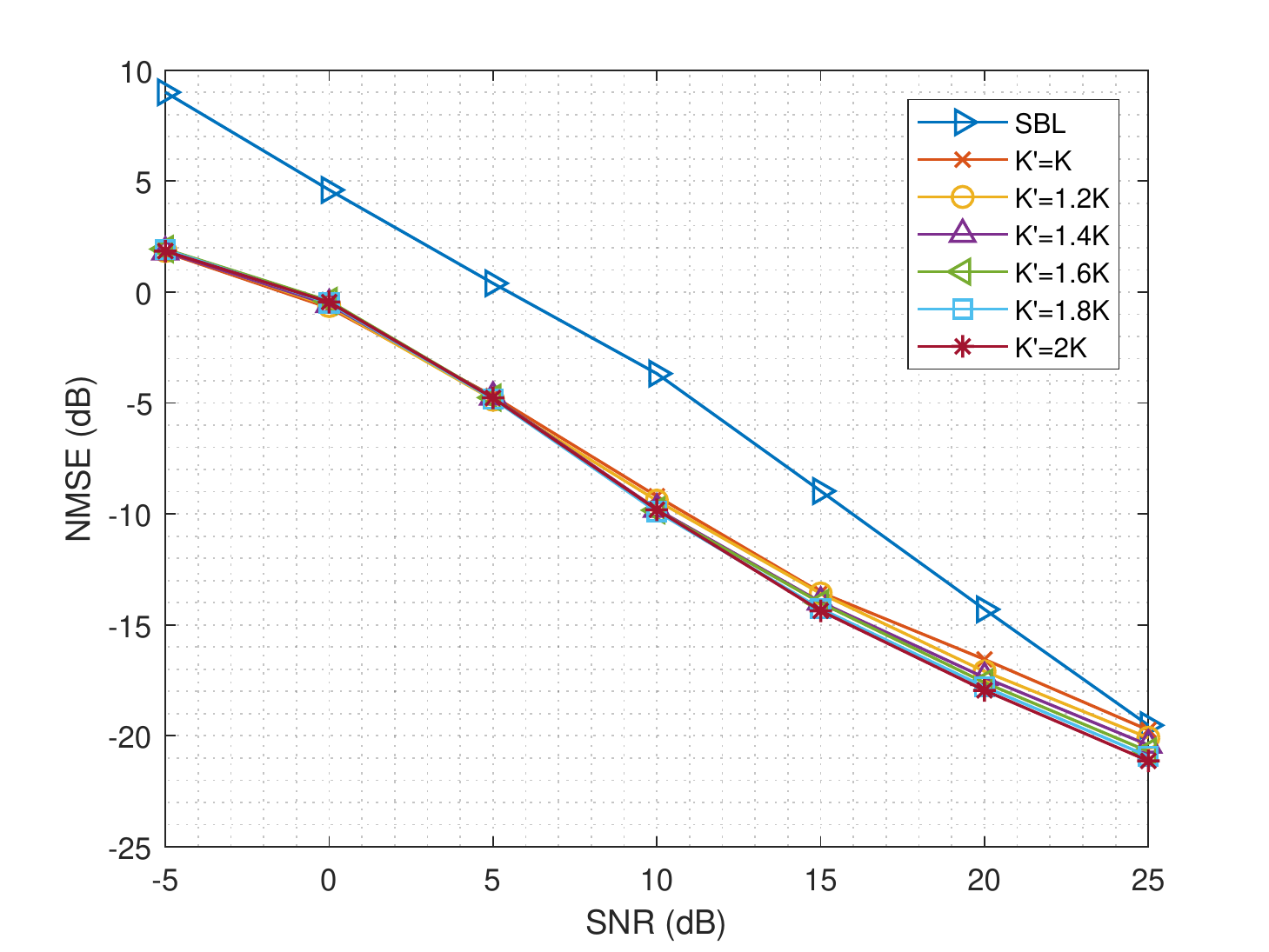}
    \caption{}
    \label{GDSvsEMSvarthetaA}
    \end{subfigure}
    \begin{subfigure}{0.48\linewidth}
    \centering
    \includegraphics[width=\linewidth]{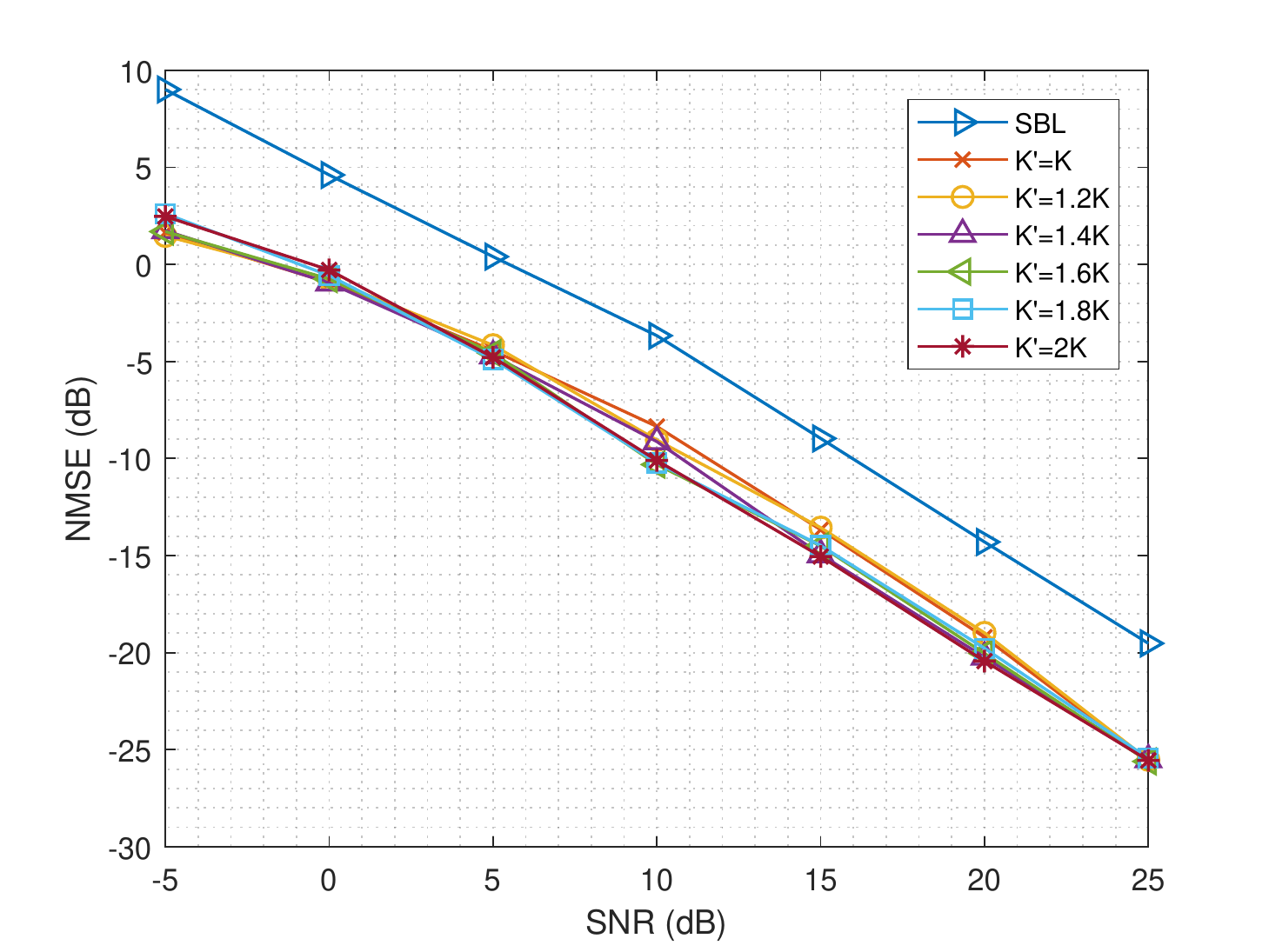}
    \caption{}
    \label{GDSvsEMSvarthetaB}
    \end{subfigure}
    \caption{NMSE performances of VSP-GD and VSP-ELBO under different ${K'}$. $N=50$, $K=10$, $L=1$, $M = 25$, and $T_{\text{out}} = 2$. The NMSE of SBL is also provided as benchmark. (a) NMSEs of VSP-GD versus the SNR under different ${K'}$. For the GD--based solver, $T_{\text{in}}=7000$. (b) NMSEs of VSP-ELBO versus the SNR under different ${K'}$. For the ELBO--based solver $T_{\text{in}}=30$. }
    \label{GDSvsEMSvartheta}
    \vspace{-0.2cm}
\end{figure*}

\begin{figure*}[t]
    \centering
    \begin{subfigure}{0.48\linewidth}
    \centering
    \includegraphics[width=\linewidth]{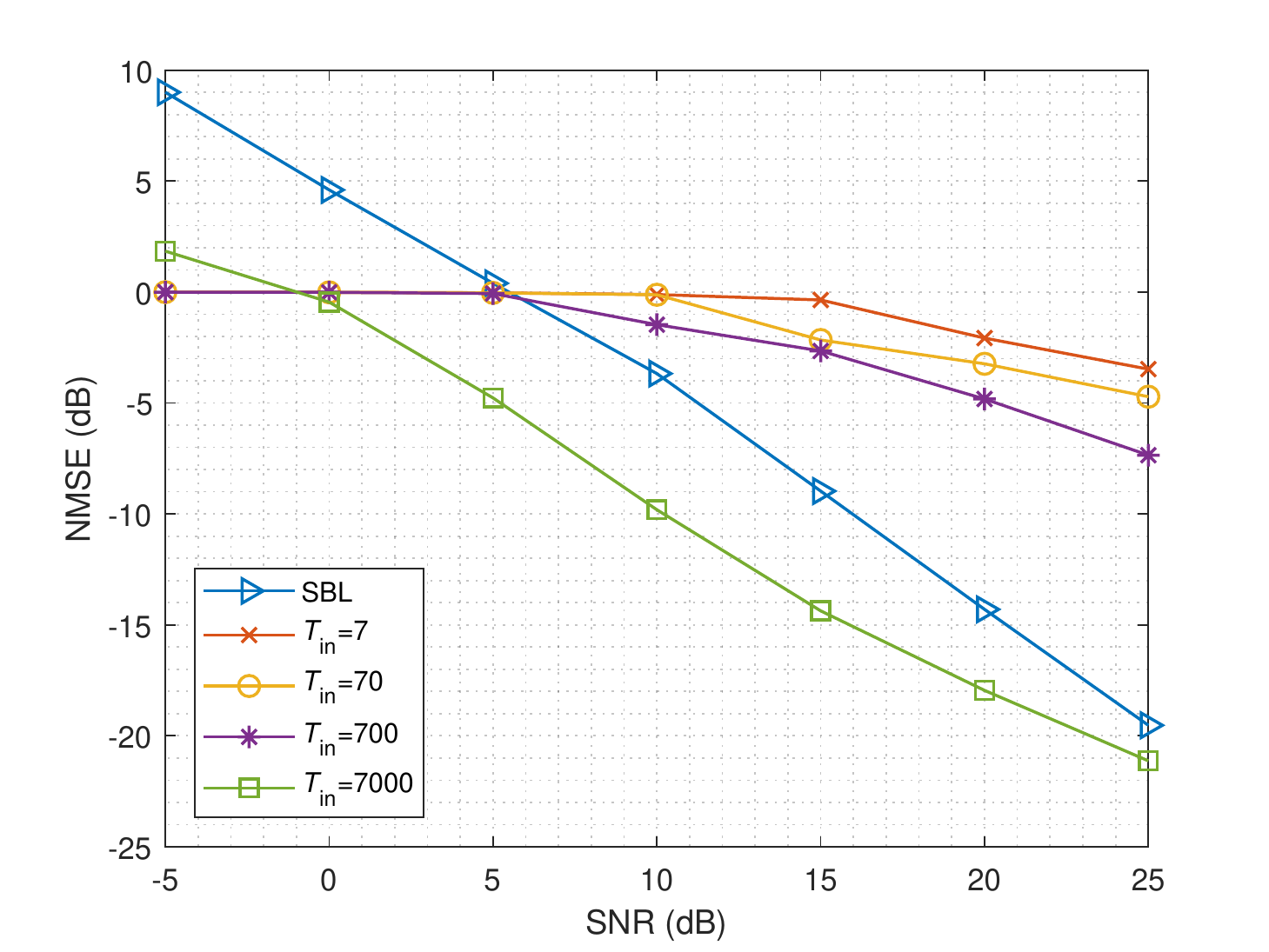}
    \caption{}
    \label{TGD}
    \end{subfigure}
    \begin{subfigure}{0.48\linewidth}
    \centering
    \includegraphics[width=\linewidth]{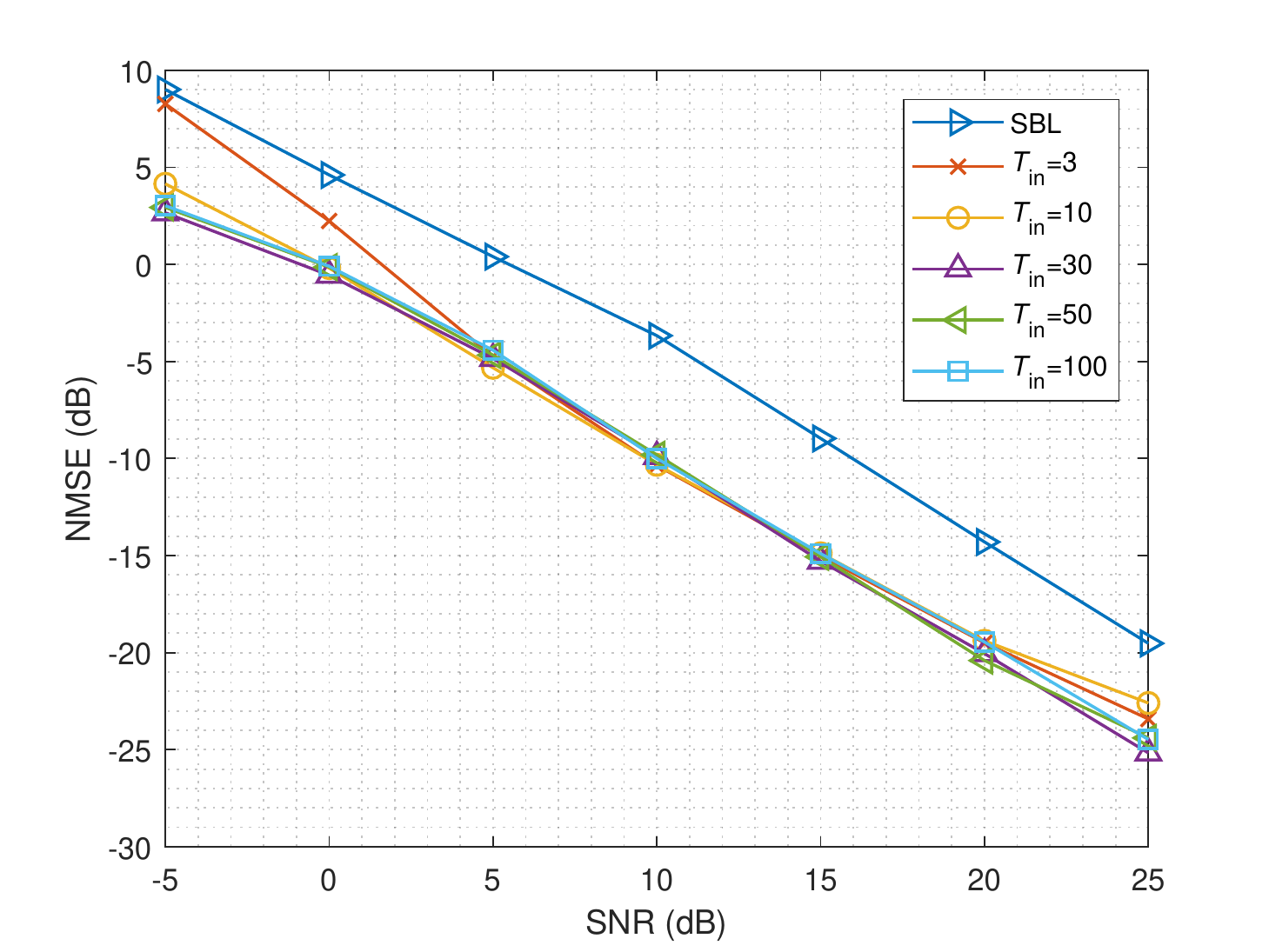}
    \caption{}
    \label{TELBO}
    \end{subfigure}
    \caption{NMSE performances of VSP-GD and VSP-ELBO under different $T_{\text{in}}$ respectively. $N=50$, $K=10$, $L=1$, $M = 25$, ${K'} = 2K$, and $T_{\text{out}} = 2$. (a) NMSEs of VSP-GD versus the SNR under different $T_{\text{in}}$. (b) NMSEs of VSP-ELBO versus the SNR under different $T_{\text{in}}$. The NMSE of SBL is also provided as a benchmark.}
    \label{TGDorELBO}
    \vspace{-0.2cm}
\end{figure*}

The overall VSP algorithm is summarized in Algorithm \ref{Algo1}. The input argument $\boldsymbol{y}$ is the noise-corrupted measurement; $\boldsymbol{A}$ is the measurement matrix; $\boldsymbol{\mu}^{(0)}$ is the initial estimate of $\boldsymbol{v}$; $a$, $b$ are the shape and the rate parameter of the Gamma distribution; $T_{\text{out}}$ is the number of the outer iteration; $T_{\text{in}}$ is the number of iterations in the GD-based and ELBO-based solvers; $\alpha$ and $\beta$ are the parameters of the Markov random field; $\rho$ is the proportion of nonzero elements in $\boldsymbol{x}$; $\vartheta$ is the coefficient; $\sigma^2$ is the variance of the Gaussian noise. In the outer iteration of VSP, the algorithm first calls the GD-based solver or the ELBO-based solver to calculate an estimate $[\mu_{g_1 \rightarrow v_1}, \ldots, \mu_{g_N \rightarrow v_N}]^T$ of $\boldsymbol{v}$. The estimate $[\mu_{g_1 \rightarrow v_1}, \ldots, \mu_{g_N \rightarrow v_N}]^T$ is then passed into the Markov random field for further processing to encourage the block sparsity of $\{\mu_{g_i \rightarrow v_i}\}$ (Lines 6 to 9 of Algorithm \ref{Algo1}). The output $\left[\mu_{v_1 \rightarrow g_1}, \ldots, \mu_{v_N \rightarrow g_N} \right]^T$ is taken as the initial value of the GD-based/ELBO-based solver in the next iteration (Line 10 of Algorithm \ref{Algo1}). In a sense, the role of the Markov random field in VSP is to iteratively adjust the initial variances of the GD-based/ELBO-based solver according to the block-sparse prior of $\boldsymbol{x}$. The final estimate of $\boldsymbol{x}$ is given by $\{\mu_{g_i \rightarrow v_i}\}$ via \eqref{m}. Since the state of the variances (i.e., $\{s_i\}$) plays a crutial role in message propagation, we refer to our proposed algorithm as variance state propagation.

\begin{figure*}[t]
    \centering
    \begin{subfigure}{0.48\linewidth}
    \centering
    \includegraphics[width=\linewidth]{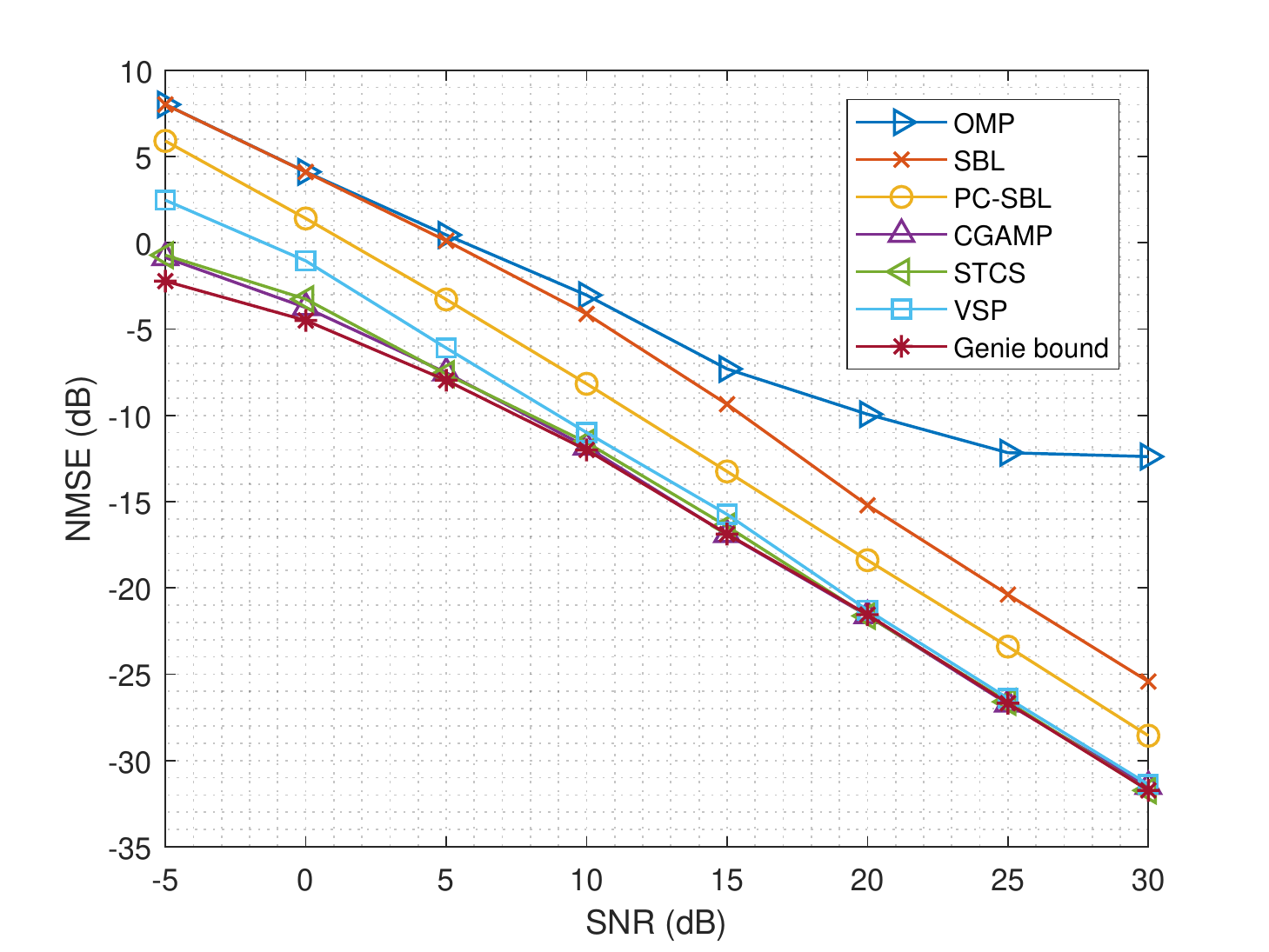}
    \caption{}
    \label{iidGaussiana}
    \end{subfigure}
    \begin{subfigure}{0.48\linewidth}
    \centering
    \includegraphics[width=\linewidth]{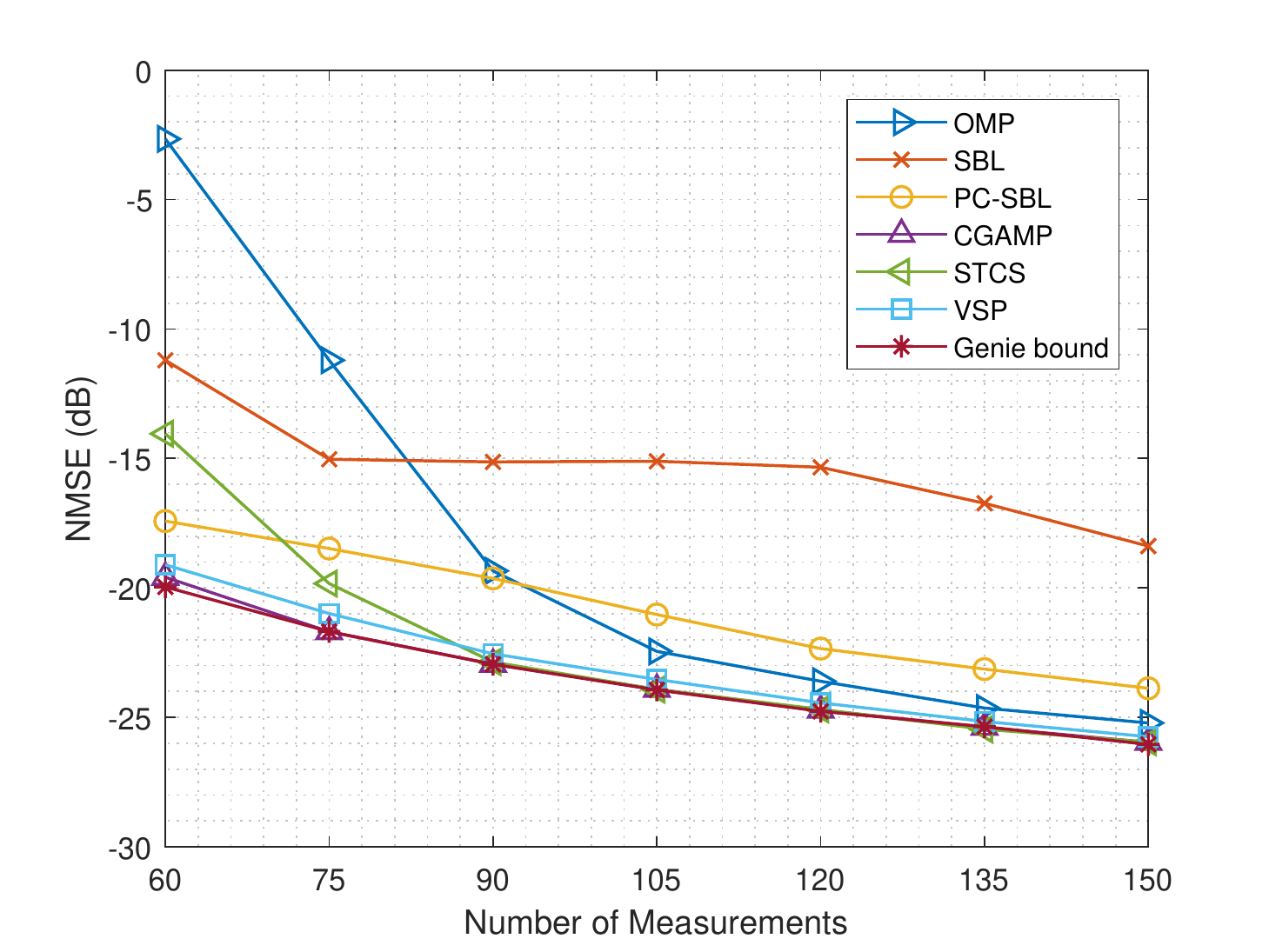}
    \caption{}
    \label{iidGaussianb}
    \end{subfigure}
    \caption{NMSEs of the respective algorithms under Gaussian measurement matrices. (a) NMSEs versus the SNR. $N=200$, $K=30$, $L=1$, and $M = 75$. (b) NMSEs versus the number of measurements. $N=200$, $K=30$, $L=1$, and $\text{SNR} = 20 ~ \text{dB}$.}
    \label{iidGaussian}
    \vspace{-0.2cm}
\end{figure*}


\begin{figure*}[t]
    \centering
    \begin{subfigure}{0.48\linewidth}
    \centering
    \includegraphics[width=\linewidth]{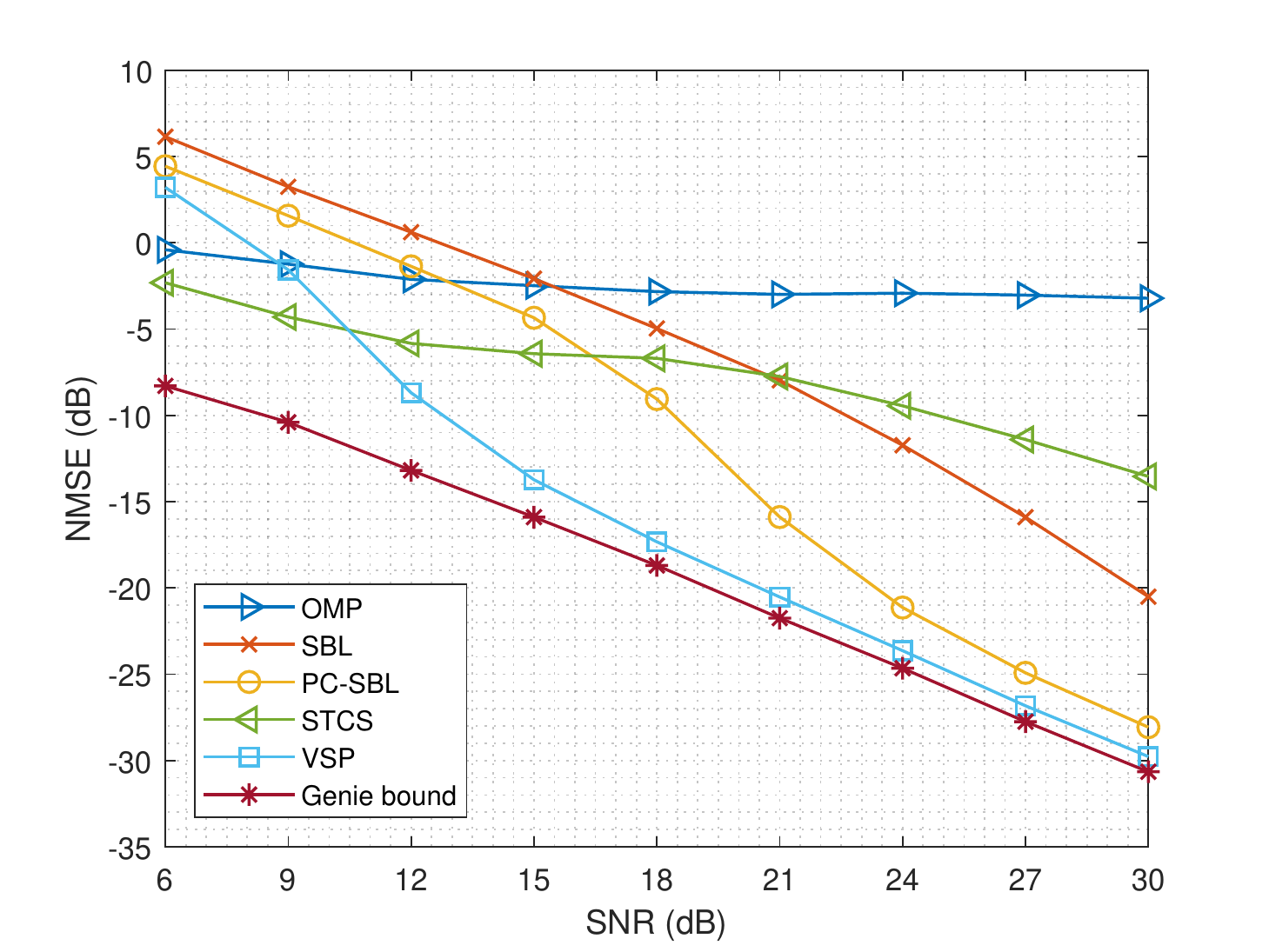}
    \caption{}
    \label{Tcrapa}
    \end{subfigure}
    \begin{subfigure}{0.48\linewidth}
    \centering
    \includegraphics[width=\linewidth]{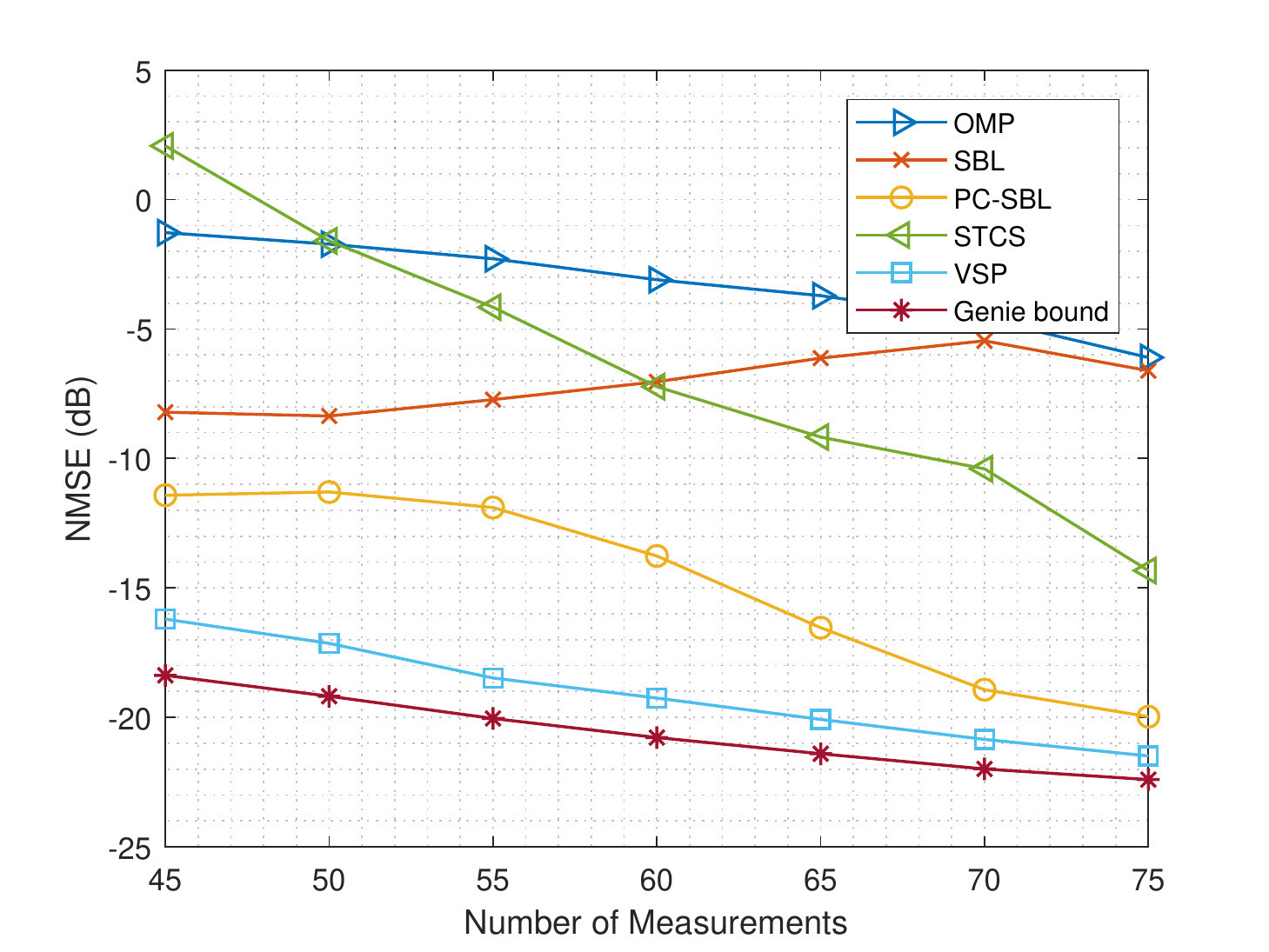}
    \caption{}
    \label{Tcrapb}
    \end{subfigure}
    \caption{NMSEs of the respective algorithms under cropped-Hermitian measurement matrices. (a) NMSEs versus the SNR. $N=100$, $K=20$, $L=2$, and $M = 60$. (b) NMSEs versus the number of measurements. $N=100$, $K=20$, $L=2$, and $\text{SNR} = 20 ~ \text{dB}$.}
    \label{Tcrap}
    \vspace{-0.2cm}
\end{figure*}

\begin{figure*}[t]
    \centering
    \begin{subfigure}{0.48\linewidth}
    \centering
    \includegraphics[width=\linewidth]{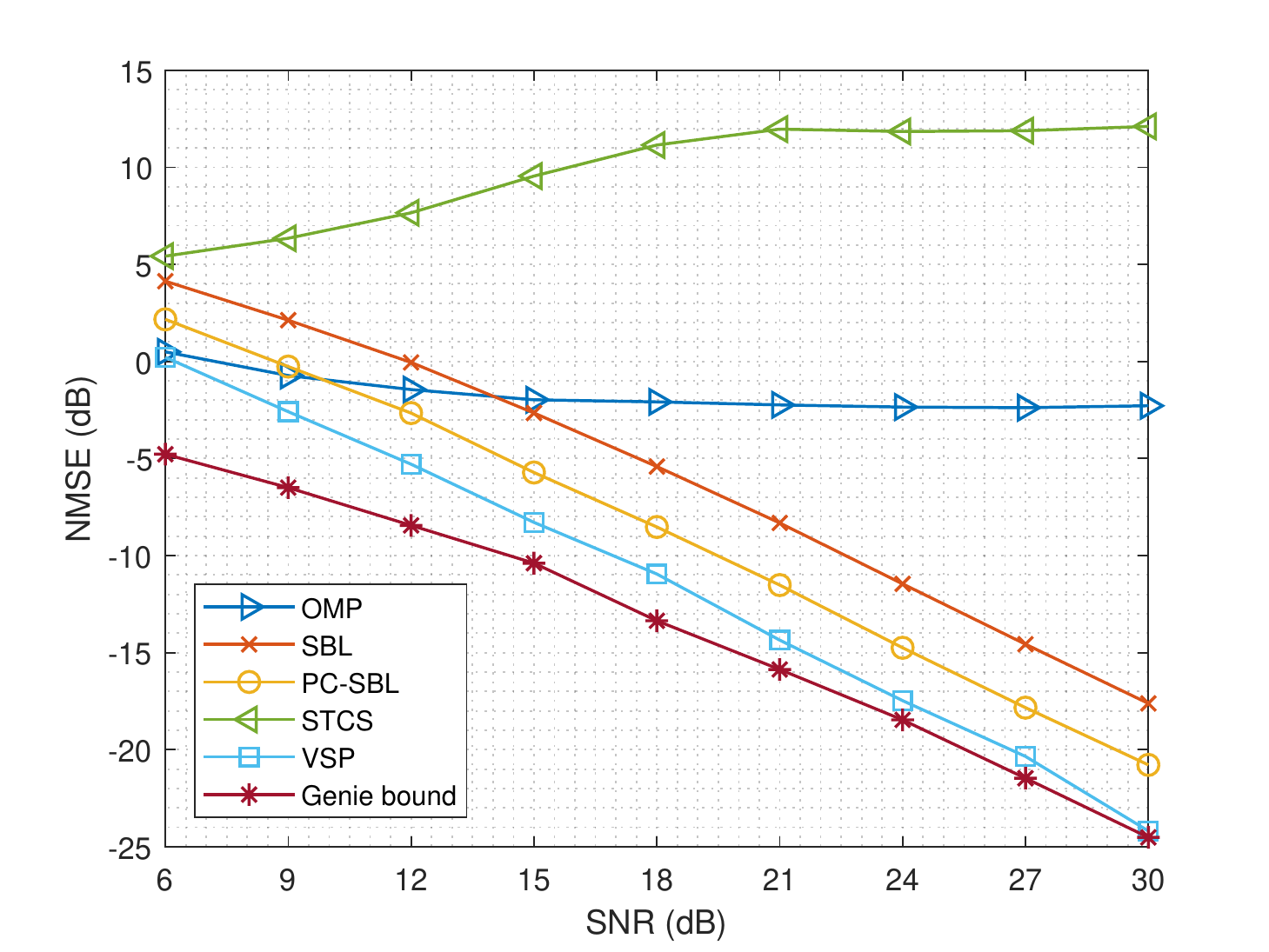}
    \caption{}
    \label{ExpGaussiana}
    \end{subfigure}
    \begin{subfigure}{0.48\linewidth}
    \centering
    \includegraphics[width=\linewidth]{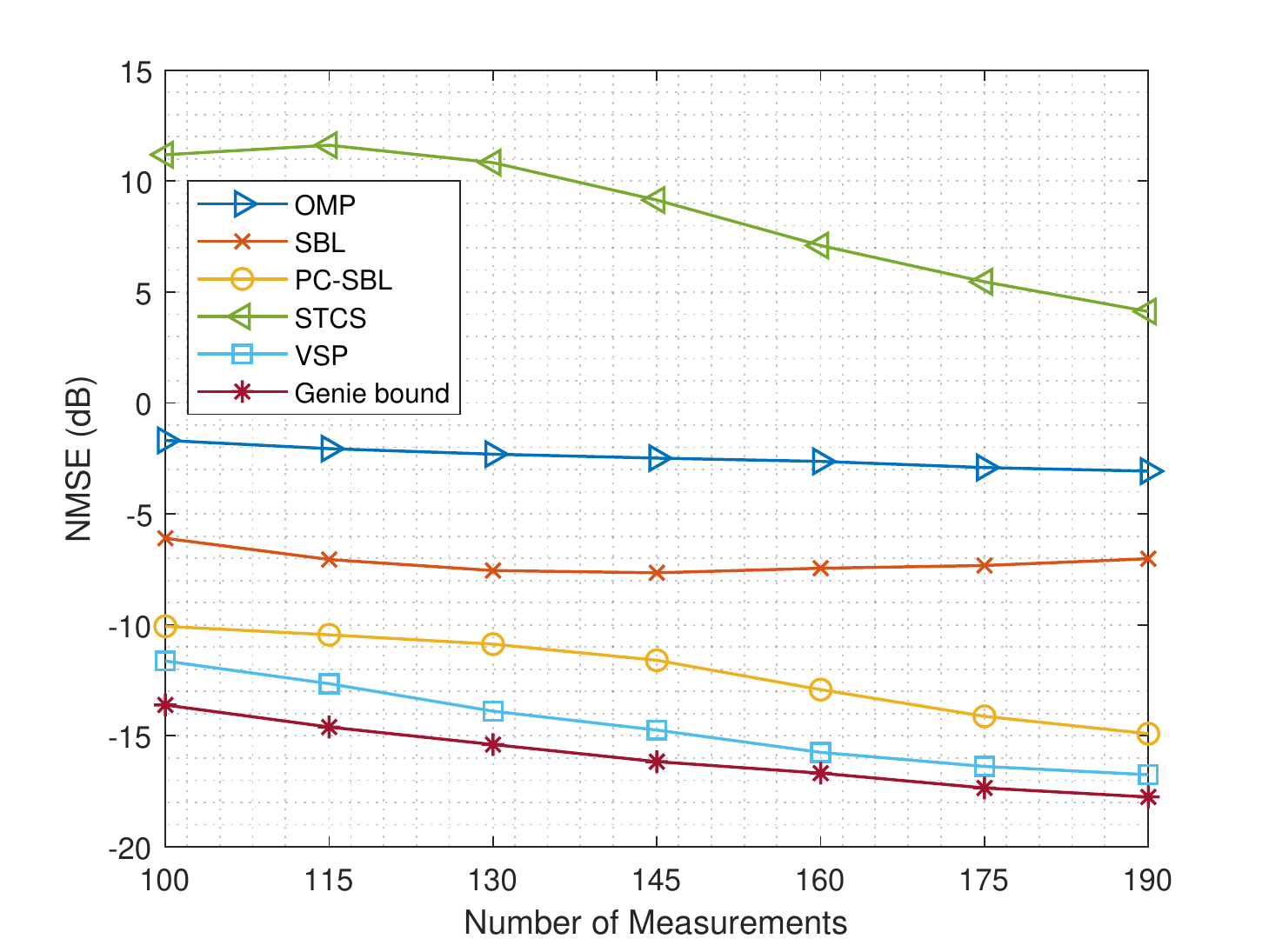}
    \caption{}
    \label{ExpGaussianb}
    \end{subfigure}
    \caption{NMSEs of respective algorithms under concatenated-exponential-Gaussian measurement matrices. (a) NMSEs versus the SNR. $N=300$, $K=50$, $L=3$, and $M = 120$. (b) NMSEs versus the number of measurements. $N=300$, $K=50$, $L=3$, and $\text{SNR} = 20 ~ \text{dB}$.}
    \label{ExpGaussian}
    \vspace{-0.2cm}
\end{figure*}

\subsection{Further Discussions}

The total complexity of the VSP algorithm consists of the implementation of Algorithm \ref{GDS} (or Algorithm \ref{EMS}) and the message passing steps of Algorithm \ref{Algo1}. The complexities of both Algorithms \ref{GDS} and \ref{EMS} are dominated by the calculation of $\boldsymbol{\Phi}$ in \eqref{phi}. According to the Woodbury matrix identity, this $N \times N$ matrix inversion can be converted to an $M \times M$ matrix inversion, which requires $\mathcal{O}(M^3)$ flops per iteration. Consequently, the complexities of Algorithms \ref{GDS} and \ref{EMS} are both $\mathcal{O}(T_{\text{in}} M^3)$. The calculation of the messages in steps 6--10 of Algorithm \ref{Algo1} requires the complexity of $\mathcal{O}(N)$. Therefore, by considering the outer iteration, the total complexity of the VSP algorithm is $\mathcal{O}(T_{\text{out}}T_{\text{in}}M^3)$.

Table \ref{table1} shows the complexity and the required prior information of VSP and some other popular compressed sensing algorithms. Compared with other methods, VSP has a significant advantage in its near-optimal performance and robustness to the measurement matrix, while its computational complexity is acceptable. VSP introduces variance variables in the probability model, resulting in a much lower correlation between messages passed on the factor graph. VSP thus inherits the superior performance of the message passing-based algorithms and maintains a good robustness to the measurement matrix. Further, by using the Markov random field to model the states of the variance variables, VSP well exploits the prior knowledge of block sparsity. The performance of VSP is examined in the next section.



\section{Numerical Results}\label{Simu}
We now carry out simulations to illustrate the performance of our proposed VSP algorithm. We first test the performance of the VSP algorithm using the GD-based solver (VSP-GD) and the VSP algorithm using the ELBO-based solver (VSP-ELBO) under a relatively simple environment, and then compare VSP with the other existing block-sparse signal recovery algorithms in several practical applications.

\subsection{GD Solver vs. ELBO Solver}\label{SecGDvsELBO}
In this subsection we compare the performance of VSP-GD and VSP-ELBO under different parameter settings.
To evaluate the recovery performance, we introduce the normalized mean square error (NMSE) metric, which is calculated by averaging normalized squared errors $||\hat{\boldsymbol{x}}-\boldsymbol{x}||_2^2 / ||\boldsymbol{x}||_2^2$ over independent trials, where $\hat{\boldsymbol{x}}$ denotes an estimate of $\boldsymbol{x}$.
In our experiments, the block-sparse signals are generated in a similar way as in \cite{fang2014pattern}. The sparse signal $\boldsymbol{x} \in \mathbb{C}^N$ contains $K$ nonzero coefficients partitioned into $L$ blocks, and the location and the size of each block are randomly assigned. The block sizes $\{B_l\}_{l=1}^L$ are determined as follows: we randomly generate $L$ positive random variables $\{r_l\}_{l=1}^L$ with their sum equal to one. Then we set $B_l=\lceil K\tau_l \rceil$ for the first $L-1$ blocks and $B_L=K-\sum_{l=1}^{L-1} B_l$ for the last block, where $\lceil x \rceil$ denotes the ceiling operator that gives the smallest integer no smaller than $x$. Similarly, we partition the $N$-dimensional vector into $L$ super-blocks using the same set of values $\{r_l\}_{l=1}^L$, and place each of the $L$ nonzero blocks into one unique super-block with a randomly generated starting position, where the starting position is carefully selected to prevent the nonzero block from going beyond the super-block. The nonzero coefficients and the elements of the measurement matrix $\{x_k\}_{k=1}^K$ are independently drawn from the standard complex Gaussian (SCG) distribution \cite{tse2005fundamentals} with zero mean and unit variance. The given results are averaged by 200 independent trails.

We first test the impact of the choice of $K'$ on the performance of VSP-GD and VSP-ELBO. In \eqref{kappa} we notice that when updating parameters of the check function $f_i = p(v_i|s_i)$, the sparsity $K$ is needed to determine ${K'}$. In practice, accurate knowledge of $K$ is a pretty strong prior that may be unavailable. In many cases, we may only know the approximate number of nonzero elements in $\boldsymbol{x}$. Therefore, we hope that the VSP algorithm is not sensitive to the value of ${K'}$. In our simulation, we test the sensitivity of VSP to ${K'}$ by fixing $K$ and adjusting $\vartheta$. The NMSEs of VSP-GD and VSP-ELBO versus the signal-noise-ratio (SNR) are presented in Fig. \ref{GDSvsEMSvartheta}. The SNR is defined as $20\log \{||\boldsymbol{Ax}||_2/\sigma \}$ in $\text{dB}$, where $\boldsymbol{A}$ is the measurement matrix in \eqref{LinearModel}, and $\sigma$ is the standard deviation of the complex Gaussian noise. Fig. \ref{GDSvsEMSvarthetaA} and Fig. \ref{GDSvsEMSvarthetaB} show the NMSE curves for VSP-GD and VSP-ELBO at different values of ${K'}$, respectively. The NMSE curve of SBL is also provided for comparision. We see that when ${K'}$ ranges from $K$ to $2K$, both VSP-GD and VSP-ELBO exhibit significant performance gains over the baseline SBL, and more importantly, the performance gains are generally not sensitive to the value of $K'$.
In Fig. \ref{GDSvsEMSvarthetaA} we observe that, when the SNR is less than 10 dB, the NMSE performance of VSP-GD is relatively insensitive to ${K'}$. When the SNR is greater than 10 dB, a larger ${K'}$ brings a slight gain.
In Fig. \ref{GDSvsEMSvarthetaB} we also observe that a larger ${K'}$ ($1.6K \sim 2K$) makes the performance of VSP-ELBO slightly better.
Comparing Fig. \ref{GDSvsEMSvarthetaA} and Fig. \ref{GDSvsEMSvarthetaB}, we find that as the SNR increases, the gap between VSP-GD and SBL narrows, whereas VSP-ELBO does not suffer from this problem.

We now examine the impact of $T_{\text{in}}$ for VSP. We choose ${K'}=2K$ and $T_{\text{out}}=2$. The block-sparse signals and the measurement matrices are generated in the same way as in the previous experiment. Fig. \ref{TGDorELBO} shows the NMSE of VSP-GD versus SNR with different $T_{\text{in}}$. The NMSE of SBL versus SNR is also provided for comparision. In Fig. \ref{TGD}, we observe that VSP-GD needs a large $T_{\text{in}}$ ($T_{\text{in}}=7000$) to ensure good performance. In contrast, VSP-ELBO achieves a significant NMSE gain over SBL at a relatively small $T_{\text{in}}$ value ($T_{\text{in}} = 10$ in Fig. \ref{TELBO}).

Through the above two sets of experiments, we find that VSP-ELBO is superior to VSP-GD in terms of both recovery performance and computational complexity. In subsequent experiments for comparision with the other existing methods, ``VSP'' always refers to ``VSP-ELBO'' with $T_{\text{out}}=2$, $T_{\text{in}}=30$, and ${K'}=2K$.

\subsection{Synthetic Data}\label{SecOneDim}
In this subsection we evaluate the recovery performance of the VSP for synthetic block-sparse signals. We consider three different measurement matrix structures to test the robustness of the VSP algorithm. Here, the block-sparse signals are generated in the same manner as described in Section \ref{SecGDvsELBO}.
The existing algorithm for sparse signal recovery, including the orthogonal matching pursuit (OMP) \cite{tropp2007signal}, conventional sparse Bayesian learning (SBL) \cite{tipping2001sparse}, pattern-coupled sparse Bayesian learning (PC-SBL), clustered Gaussian approximate message passing (CGAMP) \cite{he2019super}, and structured turbo compressed sensing (STCS) \cite{chen2017structured} are taken into account for comparison.


\begin{figure*}[t]
    \centering
    \includegraphics[width=0.24\linewidth]{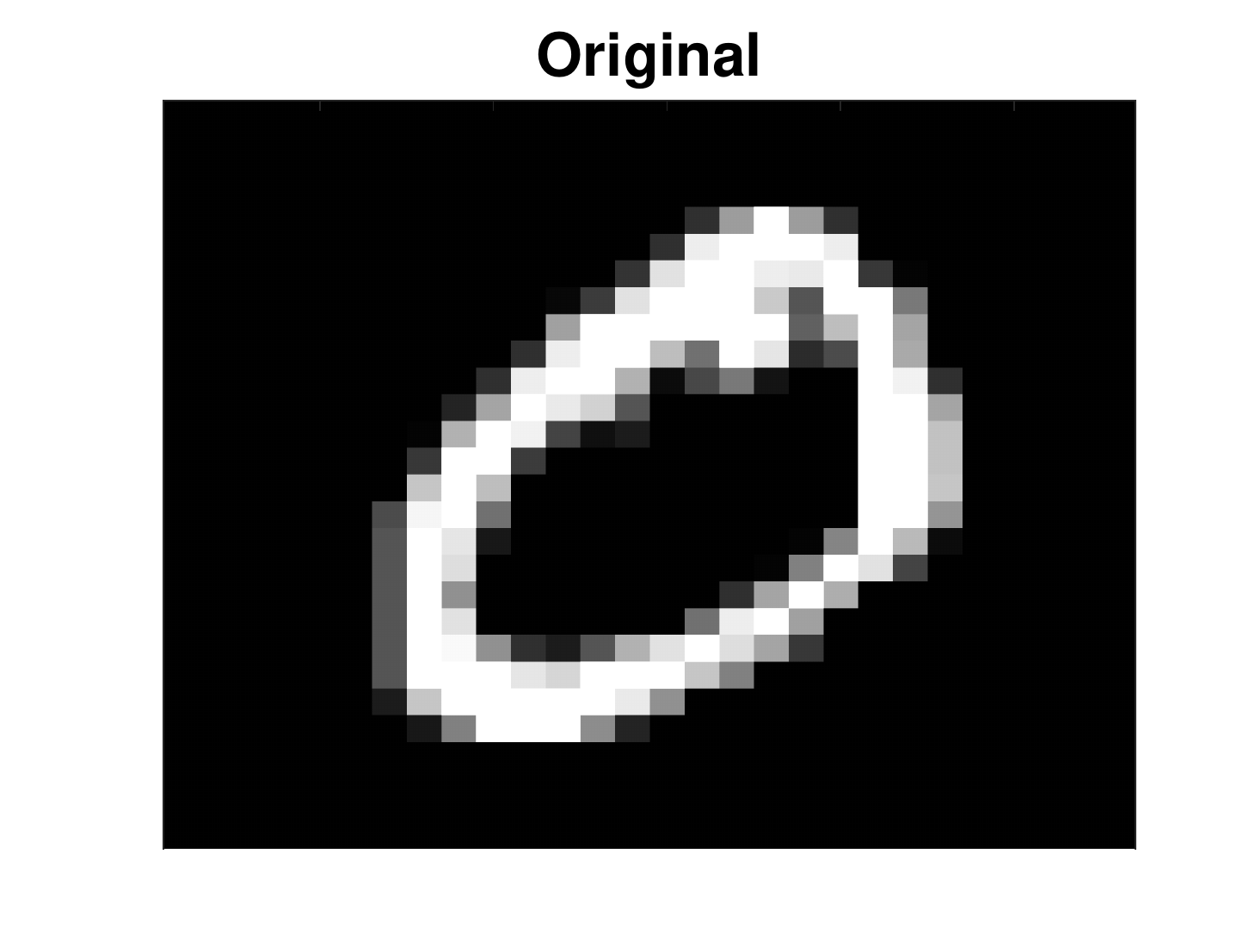}
    \includegraphics[width=0.24\linewidth]{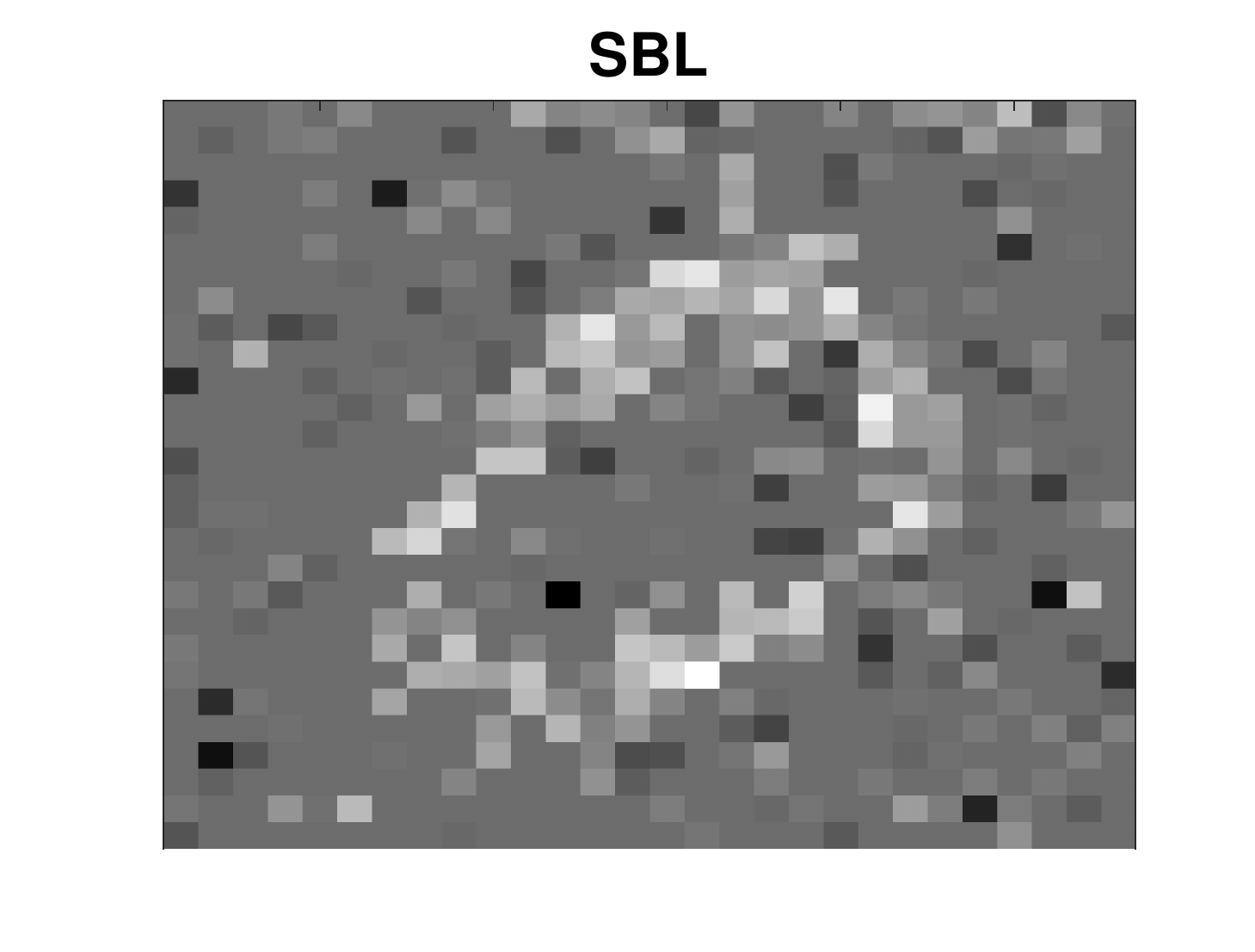}
    \includegraphics[width=0.24\linewidth]{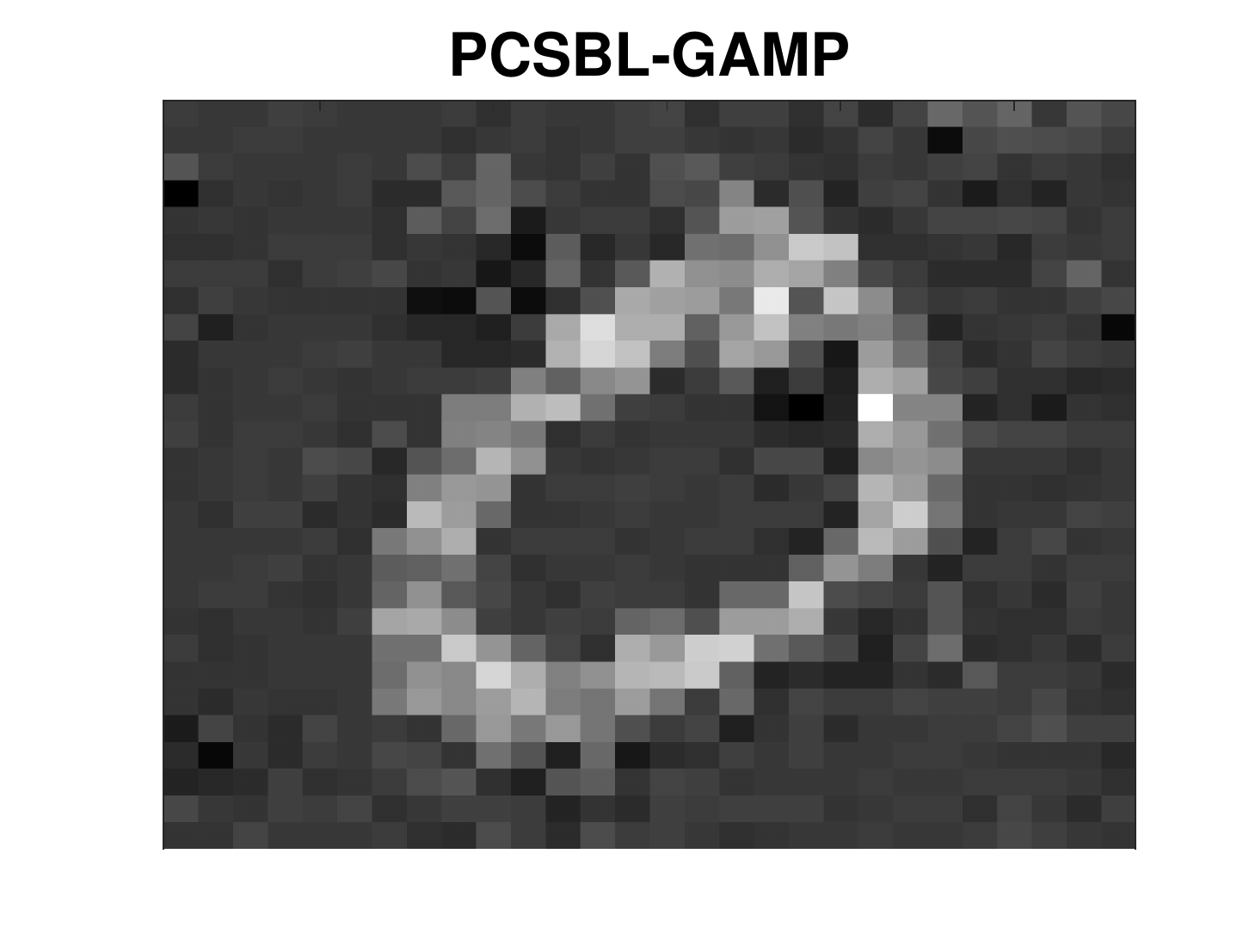}
    \includegraphics[width=0.24\linewidth]{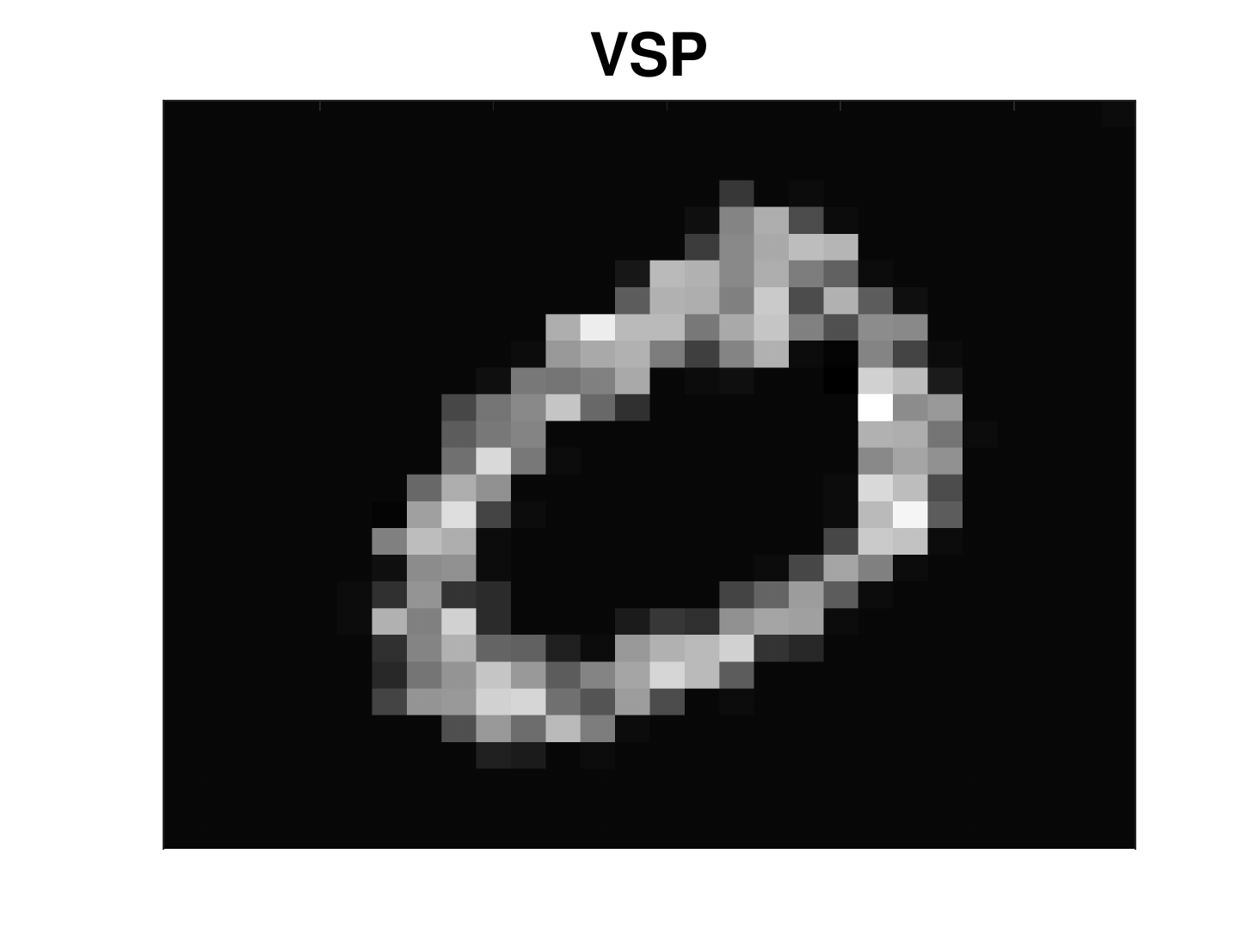}
    \caption{Original image of handwritten digit ``0'' and the reconstructed images by SBL, PCSBL-GAMP, and VSP under the Gaussian measurement matrix. $\text{SNR} = 10 ~ \text{dB}$.}
    \label{Letter0}
    \vspace{-0.2cm}
\end{figure*}

\begin{figure*}[t]
    \centering
    \includegraphics[width=0.24\linewidth]{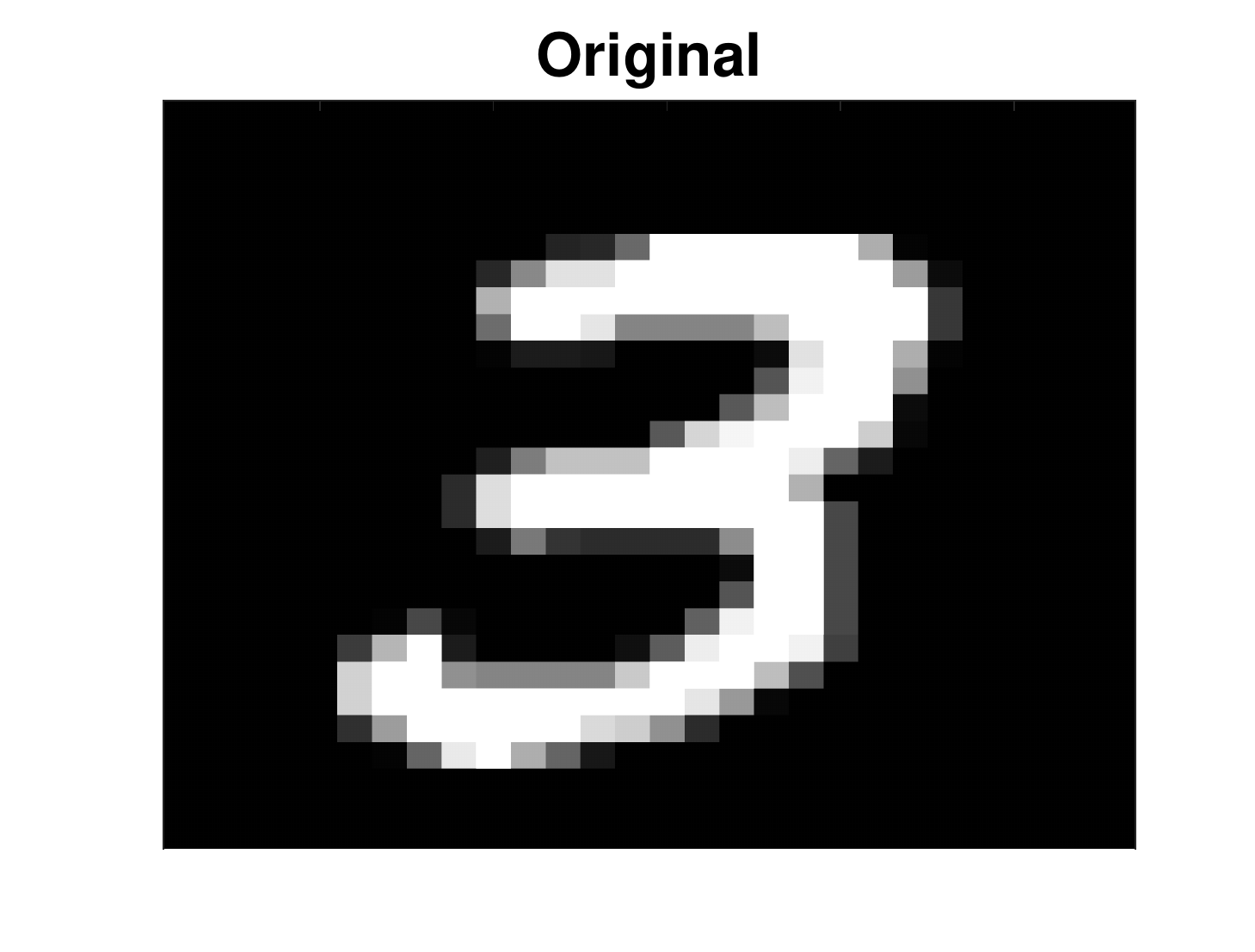}
    \includegraphics[width=0.24\linewidth]{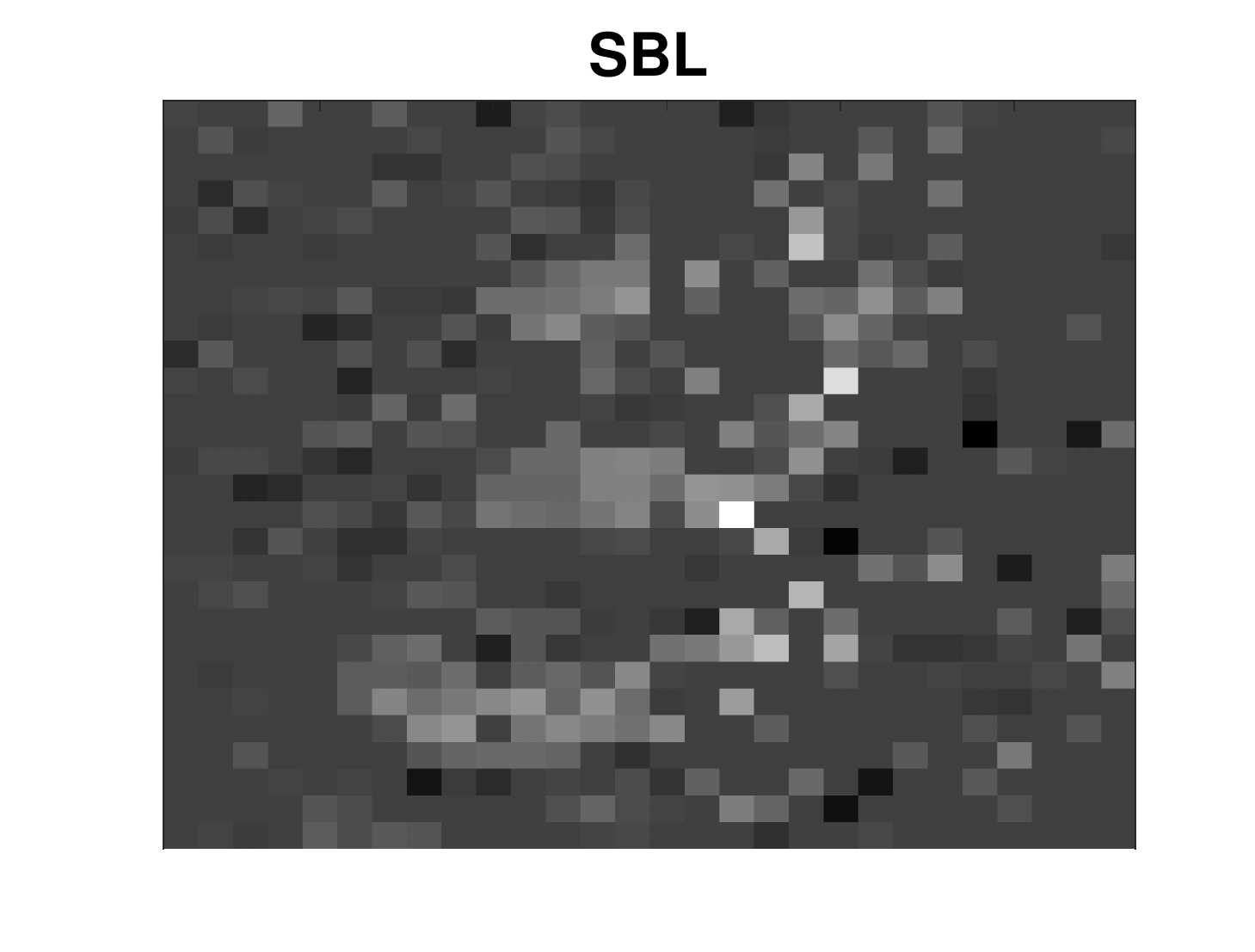}
    \includegraphics[width=0.24\linewidth]{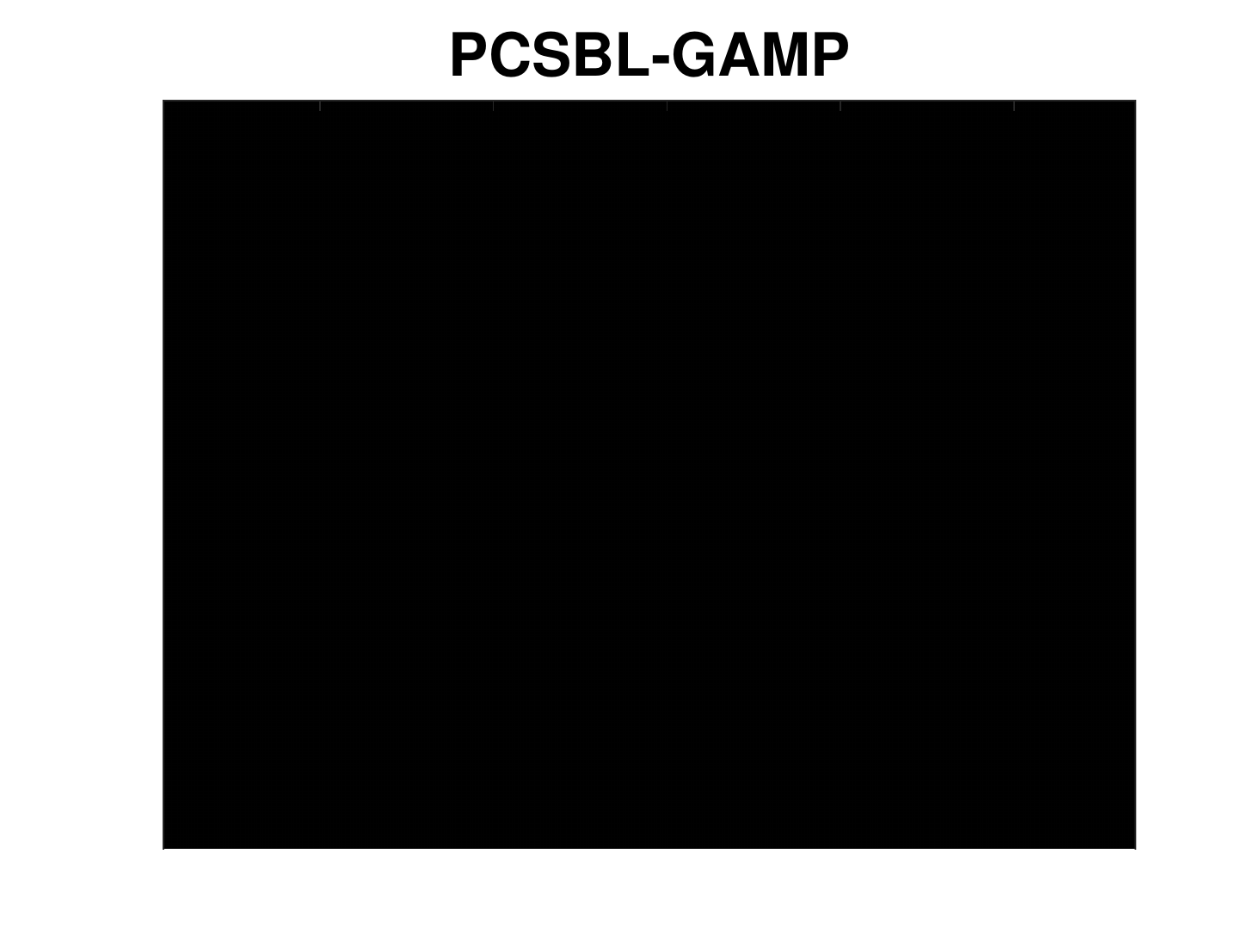}
    \includegraphics[width=0.24\linewidth]{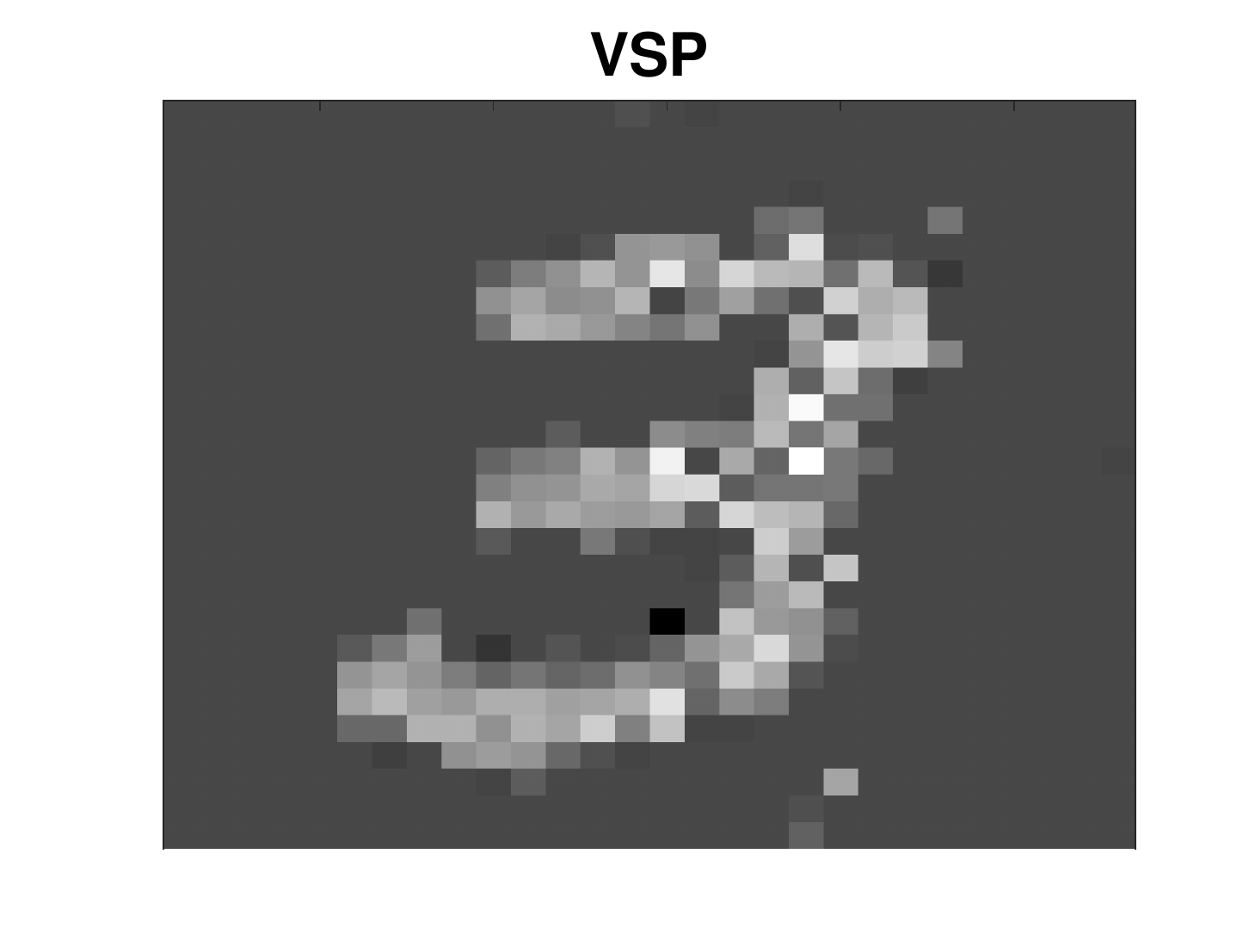}
    \caption{Original image of handwritten digit ``3'' and the reconstructed images by SBL, PCSBL-GAMP, and VSP under the concatenated-exponential measurement matrix. $\text{SNR} = 10 ~ \text{dB}$.}
    \label{Letter3}
    \vspace{-0.2cm}
\end{figure*}

The NMSEs of the respective algorithms versus the SNR and the number of measurements are depicted in Fig. \ref{iidGaussian}, Fig. \ref{Tcrap}, and Fig. \ref{ExpGaussian}. The SNR is defined in the same way as in the preceding subsection. We also plot a genie bound as a benchmark, which is obtained by a linear minimum mean-square error (LMMSE) estimator with perfectly known non-zero positions of $\boldsymbol{x}$. The results are averaged by 500 realizations.

Fig. \ref{iidGaussian} is obtained under complex Gaussian measurement matrices, namely, the elements of $\boldsymbol{A}$ are independently drawn from the SCG distribution. In Fig. \ref{iidGaussiana} we observe that the NMSEs of STCS and CGAMP almost coincide with the genie bound predicted by the LMMSE. This is consistent with our expectation since $\boldsymbol{A}$ here is a right-rotationally invariant (RRI) matrix. The excellent performance of the message passing algorithms under this scenario has been previously confirmed in \cite{chen2017structured,he2019super}.
From Fig. \ref{iidGaussiana}, we observe that although the performance (in NMSE) of VSP is not as good as CGAMP and STCS at low SNR, the NMSE of VSP can asymptotically approach the genie bound as the SNR increases. At the same time, VSP performs better than OMP, SBL, and PC-SBL throughout the entire observation range. 
In Fig. \ref{iidGaussianb}, we note that
the NMSE of CGAMP almost coincides with the genie bound. STCS has a similar performance with CGAMP when a large number of measurements is avaiable, but exhibits instability when the number of measurements is less than $90$. VSP performs significantly better than SBL and PC-SBL in the entire observation range, and its gap from the genie bound is always kept small (within $1 ~ \text{dB}$).

Fig. \ref{Tcrap} is obtained under cropped-Hermitian measurement matrices, namely, the measurement matrix $\boldsymbol{A} \in \mathbb{C}^{M \times N}$ in each independent trial is generated in the following manner. First we generate a square matrix $\boldsymbol{A}_1 \in \mathbb{C}^{N \times N}$ from the SCG distribution and accordingly form a Hermitian matrix $\boldsymbol{A}_2 = \boldsymbol{A}_1 \times \boldsymbol{A}_1^H$. The measurement matrix $\boldsymbol{A}$ consists of the first $M$ rows of $\boldsymbol{A}_2$. Under this setting $\boldsymbol{A}$ is not a RRI matrix. In Fig. \ref{Tcrapa} we observe that, the performance of STCS deteriorates seriously compared to that in Fig. \ref{iidGaussiana}. The corresponding NMSE can not approach the genie bound any more.
CGAMP performs even worse than STCS. The NMSE of CGAMP is not given in Fig. \ref{Tcrap}, since otherwise it will make the other curves indistinguishable. It is seen that the SBL-based compressed sensing algorithms still work well under this measurement matrix, and the proposed VSP is clearly the best among them.
As the $\text{SNR}$ increases, the NMSE curve of the VSP gradually approaches the genie bound. When the $\text{SNR}$ is $30 ~ \text{dB}$, the NMSE gap between VSP and genie bound is within $1 \text{dB}$.
In Fig. \ref{Tcrapb}, we observe that the NMSE curve of VSP decreases smoothly as the number of measurements increases. In the entire observation range, 
VSP outperforms the other algorithms by a substantial margin.

Fig. \ref{ExpGaussian} is obtained under concatenated-exponential-Gaussian measurement matrices: In each independent trial, the measurement matrix $\boldsymbol{A} \in \mathbb{C}^{M \times N}$ is a concatenation of two matrices $\boldsymbol{A}_3 \in \mathbb{C}^{M \times N/2}$ and $\boldsymbol{A}_4 \in \mathbb{C}^{M \times N/2}$, i.e., $\boldsymbol{A} = \left[ \boldsymbol{A}_3 ~ \boldsymbol{A}_4 \right]$. Each element in $\boldsymbol{A}_3$ is randomly drawn from the SCG distribution, and the real part and the imaginary part of each element in $\boldsymbol{A}_4$ are randomly drawn from an exponential distribution with the rate $=3$. A measurement matrix with such unevenly distributed energy is very unfriendly to message passing based algorithms. In Fig. \ref{ExpGaussian}, we see that the STCS does not work well under this circumstance. The performance of CGAMP is omitted for the same reason as in Fig. \ref{Tcrap}. We observe that VSP again surpasses the other algorithms in terms of both recovery ability and the amount of measurements required.

\begin{figure*}[t]
    \centering
    \includegraphics[width=0.3\linewidth]{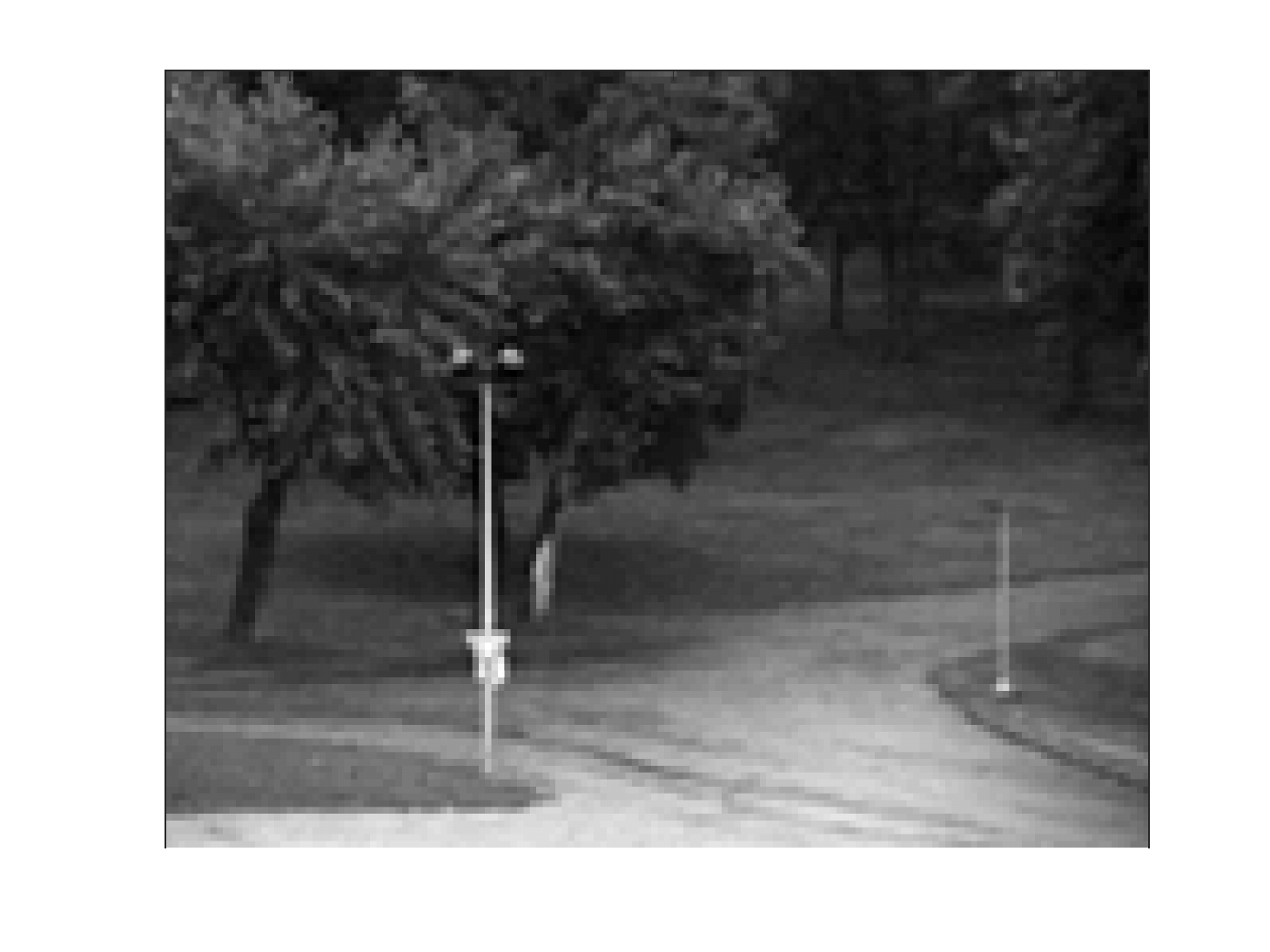}
    \includegraphics[width=0.3\linewidth]{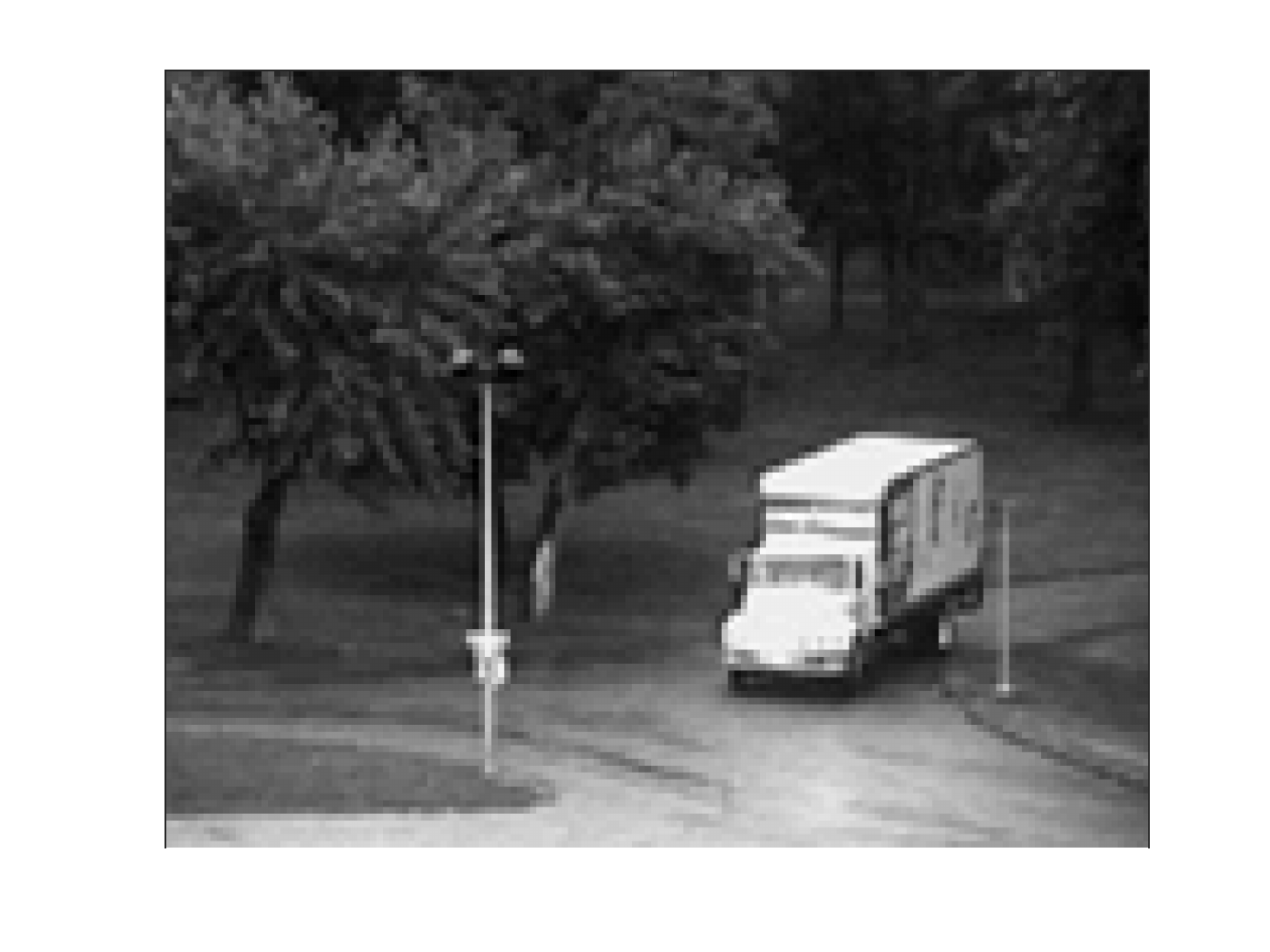}
    \includegraphics[width=0.3\linewidth]{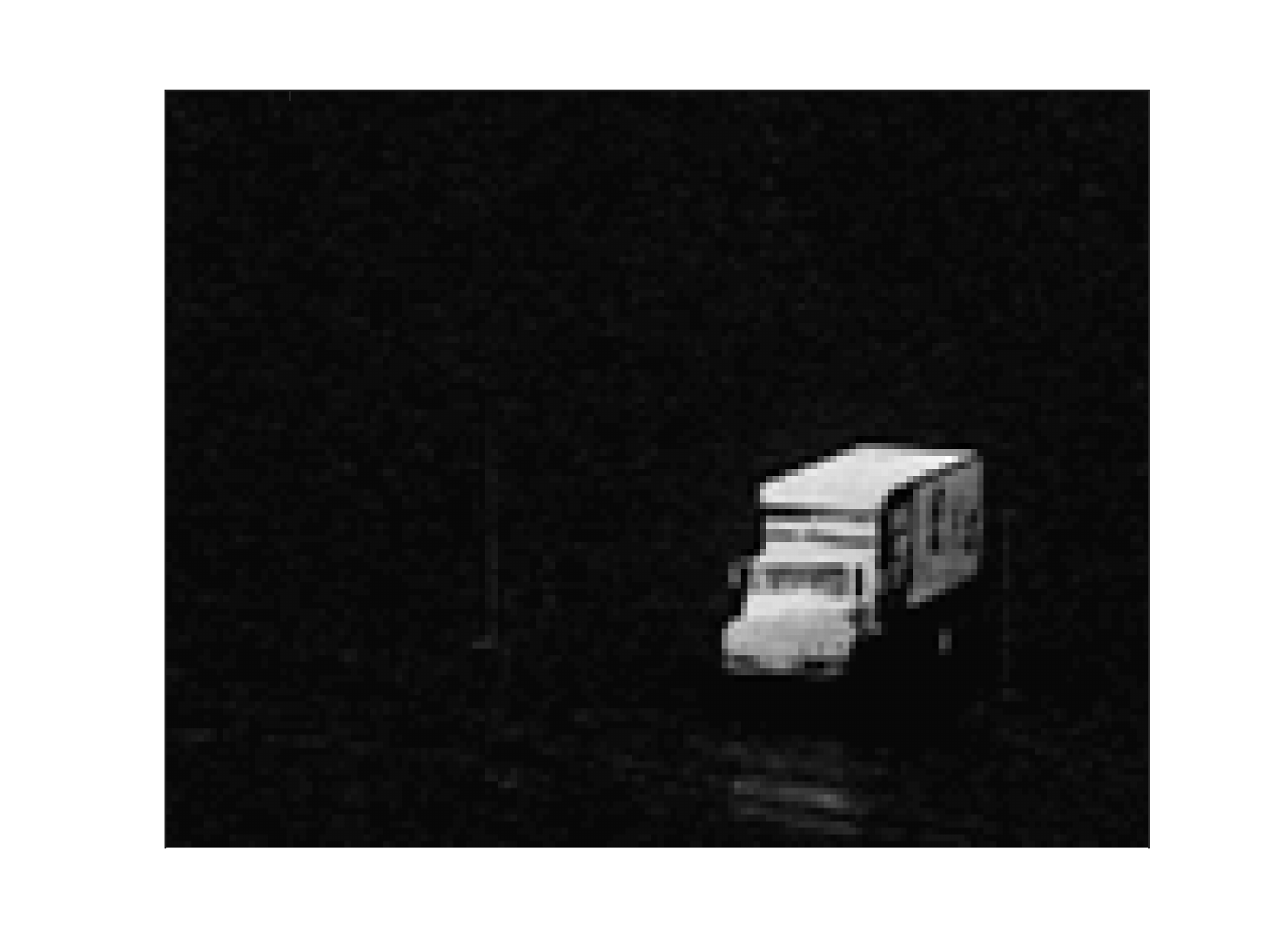}
    \includegraphics[width=0.3\linewidth]{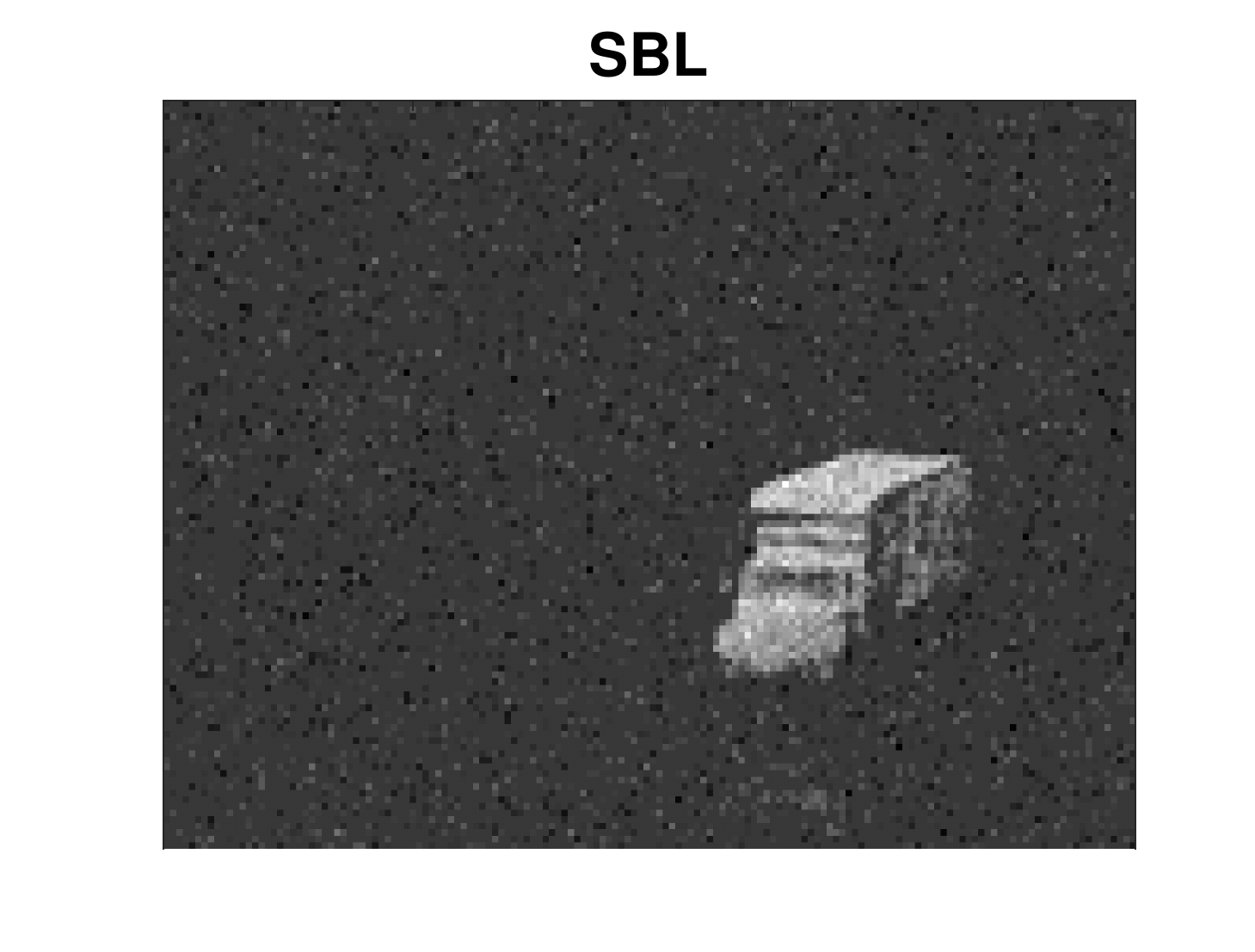}
    \includegraphics[width=0.3\linewidth]{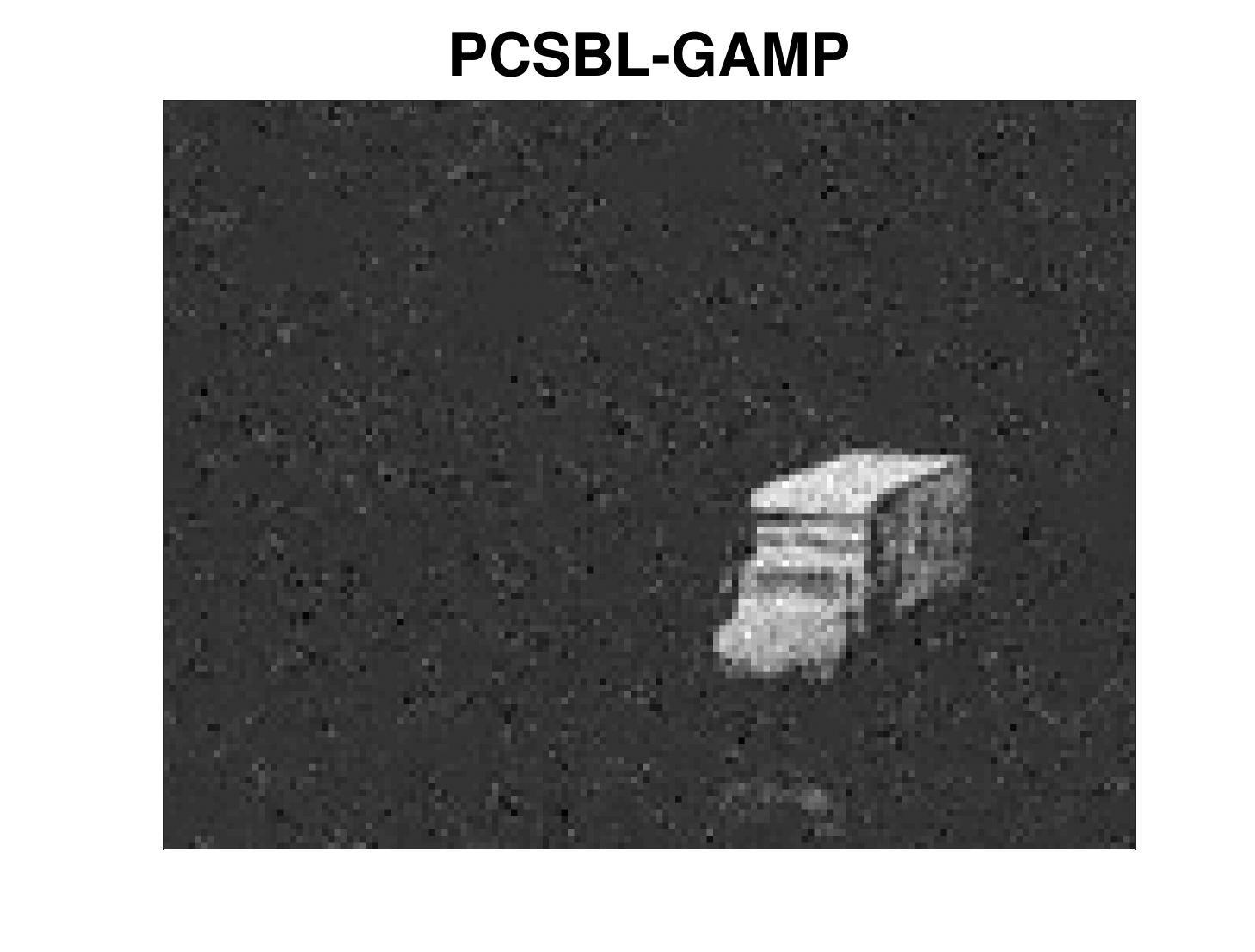}
    \includegraphics[width=0.3\linewidth]{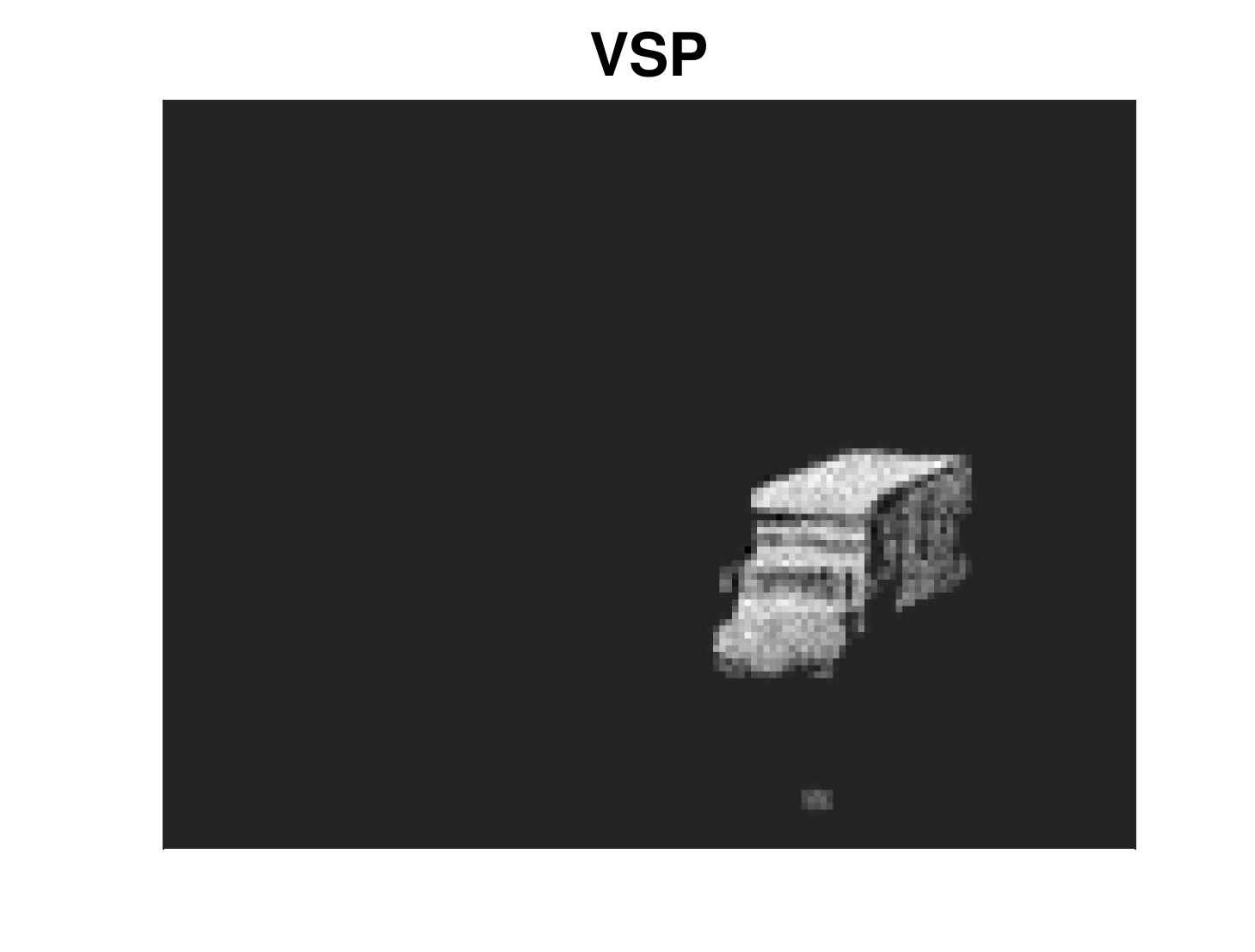}
    \caption{Top line from left to right: the background image $\boldsymbol{x}_b$ (the 70-th frame of the UCSD-rain background subtraction data set), the test image $\boldsymbol{x}_t$ (the 86-th frame of the UCSD-rain background subtraction date set), the foreground image $\boldsymbol{x}_f$. Bottom line from left to right: foreground images reconstructed by SBL, PCSBL-GAMP, VSP-SBL under concatenated-exponential measurement matrix. $\text{SNR} = 10 ~ \text{dB}$.}
    \label{Subtraction}
    \vspace{-0.2cm}
\end{figure*}

\subsection{Handwritten Digits Image Recovery}

We now test the proposed VSP algorithm on two-dimensional block-sparse signals. We carry out experiments on two handwritten digit images ($28 \times 28$ pixels) drawn from the MNIST data set \cite{lecun1998gradient}. Digit ``0'' with 176 ($22.45\%$ of total pixels) nonzero pixels and digit ``3'' with 200 ($25.51\%$ of total pixels) nonzero pixels are the $2\text{nd}$ and $8\text{th}$ samples in the MNIST training set, respectively. The gray values of each image are normalized to a range of $[0,1]$. Most of the pixels in the image are zeros and the nonzero coefficients exhibit irregular block patterns. The compressed measurements are corrupted by an additive i.i.d. Gaussian noise, i.e., $\boldsymbol{y}=\boldsymbol{Ax}+\boldsymbol{w}$, where the image is represented as a one-dimensional vector $\boldsymbol{x}$. Here we compare the recovery performance of the proposed VSP with those of SBL and PCSBL-GAMP. The PCSBL-GAMP algorithm is a generalization of the PC-SBL for two-dimensional block-sparse signals and uses Gaussian approximate message passing techniques to reduce computational complexity. In our simulations we set $M=400$. The SNR is set to $10 ~ \text{dB}$. Fig. \ref{Letter0} and Fig. \ref{Letter3} depict the original images and the reconstructed images under two different settings of the measurement matrix, respectively. Fig. \ref{Letter0} is obtained under a Gaussian measurement matrix, in which the elements are randomly drawn from a normal distribution. It can be observed that the proposed VSP algorithm provides the best visual quality with recognizable digit. The digit reconstructed by the SBL has a poor quality and can not even distinguish the boundary of the digit. The PC-SBL gives a clear boundary, but does not eliminate the noise well. The VSP not only recovers the boundary sharply, but also significantly suppresses the noise. Fig. \ref{Letter3} is obtained under a concatenated-exponential measurement matrix that is a concatenation of two matrices $\boldsymbol{A}_5$ and $\boldsymbol{A}_6$. Elements of $\boldsymbol{A}_5$ and $\boldsymbol{A}_6$ are randomly drawn from two exponential distributions with the rates $= 3$ and $= 1$, respectively. In Fig. \ref{Letter3}, it is seen that the PCSBL-GAMP totally fails due to the sensitivity of the GAMP algorithm to the measurement matrix structure. We see that in both cases, the proposed VSP offers a clearly better image recovery quality as compared with the other methods.

\subsection{Background Subtraction}

Background subtraction, also known as foreground detection,
is a technique used to automatically detect and track moving objects in videos from static cameras. Usually, the foreground interests are sparse in the spatial image domain. By exploiting this sparsity, the sparse foreground interests within a scene can be reconstructed by using compressed measurements, which improves the real-time performance of signal processing. Specifically, the idea is to reconstruct the foreground image from the noisy corrupted difference between the compressed measurements of the background image and the compressed measurements of the test image
\begin{equation}
    \mathbf{y}_f = \mathbf{y}_t - \mathbf{y}_b + \mathbf{w} = \mathbf{A}(\mathbf{x}_t - \mathbf{x}_b) + \mathbf{w} = \mathbf{A}\mathbf{x}_f + \mathbf{w}
\end{equation}
where $\mathbf{x}_t$ and $\mathbf{x}_b$ represent the test and background images, respectively; $\mathbf{y}_t$ and $\mathbf{y}_b$ denote the compressed measurements of the test and background images, respectively; $\mathbf{w}$ is the additive Gaussian noise; and $\mathbf{x}_f$ is the foreground image to be recovered. In our experiments, we use the UCSD background subtraction data set  \cite{10.1109/TPAMI.2009.112}. The data set consists of 18 video sequences collected by static cameras. We choose the 70-th frame and the 86-th frame of the ``rain'' subset as the background image $\mathbf{x}_b$ and the test image $\mathbf{x}_f$, respectively. The background image, the test image, and the foreground image are shown in the top line of Fig. \ref{Subtraction}. The foreground image is regarded as the groundtruth image. This foreground image, however, does not have a pure background since $\mathbf{x}_f = \mathbf{x}_t - \mathbf{x}_b$ is not an exactly sparse signal and contains many small nonzero components. In our experiments, the original images of $228 \times 308$ pixels are resized to $114 \times 154$ pixels. For the resized foreground image,
we have a total number of $3294$ coefficients ($18.76\%$ of total pixels) whose magnitudes are greater than $10^{-2}$. Images reconstructed by the SBL, the PC-SBL, and the VSP are depicted in the bottom line of Fig. \ref{Subtraction},  where $M=8000$ and $ \text{SNR} = 10 ~ \text{dB}$. The measurement matrix $\boldsymbol{A}$ is randomly generated with each entry independently drawn from a normal distribution. We see that our proposed PCSBL-GAMP method provides the best image quality with a clear appearance of the vehicle, whereas the object silhouettes recovered by the other methods are seriously disturbed by noise.

\section{Conclusion}\label{Conclusion}
In this paper, we developed a new sparse Bayesian learning method for recovery of block-sparse signals. A novel hierarchical Gaussian prior was proposed to characterize the block-sparse patterns of the unknown signals. The core idea of our algorithm is to iteratively update the variances in the prior Gaussian distributions. A Markov random field is combined to model the state variables of the variances of the independent Gaussian distributions. The proposed MRF-combined hierarchical model is effective and flexible to cope with various kinds of block-sparse structures. Our algorithm was developed based on the message passing principle, where for messages that are difficult to calculate, we have designed reasonable methods to achieve approximate calculations. In addition, hyperparameters can be updated within the iterative process. Simulation results show that our proposed algorithm demonstrates a superior performance over the other existing popular methods for block-sparse signal recovery.

\appendices
\section{Gradient Calculation}\label{apdixa}
The partial derivative of $\chi(\boldsymbol{y},\boldsymbol{v})$ in \eqref{chi} w.r.t. $v_i$ is given by
\begin{equation}\label{Derivative}
    \frac{\partial \chi}{\partial v_i} = - \frac{\partial \boldsymbol{m}^{H} \boldsymbol{\Phi}^{-1} \boldsymbol{m}}{\partial v_i} - \frac{\partial \ln |\boldsymbol{\Phi}|}{\partial v_i} + \frac{\partial \ln v_i}{v_i}.
\end{equation}
Furthermore,
\begin{equation}\label{p1term}
    \frac{\partial \boldsymbol{m}^{H} \boldsymbol{\Phi}^{-1} \boldsymbol{m}}{\partial v_i} = \sigma^{-4} \boldsymbol{y}^{H} \boldsymbol{A} \frac{\partial \boldsymbol{\Phi}}{\partial v_i} \boldsymbol{A}^{H} \boldsymbol{y}
\end{equation}
and
\begin{align}
    \frac{\partial \boldsymbol{\Phi}}{\partial v_i} & = \frac{\partial \left({\boldsymbol{D}}^{-1}+\sigma^{-2}\boldsymbol{A}^{H}\boldsymbol{A} \right)^{-1}}{\partial v_i} \notag \\
    & = - \boldsymbol{\Phi} \frac{\partial \left( {\boldsymbol{D}}^{-1} + \sigma^{-2}\boldsymbol{A}^{H} \boldsymbol{A} \right)}{\partial v_i} \boldsymbol{\Phi} \notag \\
    & = \boldsymbol{\Phi} \boldsymbol{E}_i \boldsymbol{\Phi}. \label{pPhipvi}
\end{align}
$\frac{\partial \ln |\boldsymbol{\Phi}|}{\partial v_i}$ can be calculated by
\begin{align}
    \frac{\partial \ln |\boldsymbol{\Phi}|}{\partial v_i}
    & = \frac{1}{|\boldsymbol{\Phi}|} \cdot \frac{\partial |\boldsymbol{\Phi}|}{\partial v_i} \notag \\
    & = \text{Tr}\left[\boldsymbol{\Phi}^{-1}\frac{\partial \boldsymbol{\Phi}}{\partial v_i} \right]. \label{pdetpvi}
\end{align}
Then, by plugging \eqref{p1term}--\eqref{pdetpvi} into \eqref{Derivative}, we obtain \eqref{pchi}.

\section{Proof of Proposition \ref{prop1}}\label{apdixb}
From \eqref{L}, we obtain
\begin{equation}
    \mathcal{L}(\mu_{g_i \rightarrow v_i}, q  (\boldsymbol{x})) \geq \mathcal{L}(\mu_{v_i \rightarrow g_i}  , q  (\boldsymbol{x})),
\end{equation}
where $q(\boldsymbol{x})$ is given by \eqref{tildeqx}. Since  $D_{\text{KL}}(p(\boldsymbol{x}|\boldsymbol{y},\boldsymbol{v})|_{\mathcal{S}_{\backslash i}, v_i = \mu_{g_i \rightarrow v_i} }||p(\boldsymbol{x}|\boldsymbol{y},\boldsymbol{v})|_{\mathcal{S}  })) \geq 0 $, we obtain
\begin{subequations}
\begin{align}
    \eta (\mu_{g_i \rightarrow v_i}) ={}& \mathcal{L}(\mu_{g_i \rightarrow v_i}, q  (\boldsymbol{x})) \notag \\
    & +D_{\text{KL}}(q  (\boldsymbol{x}) || p(\boldsymbol{x}|\boldsymbol{y},\boldsymbol{v})|_{\mathcal{S}_{\backslash i}, v_i = \mu_{g_i \rightarrow v_i} })) \\
    \geq{}&
    \mathcal{L}(\mu_{v_i \rightarrow g_i}  ,q  (\boldsymbol{x})) +D_{\text{KL}}(q  (\boldsymbol{x})||q  (\boldsymbol{x})) \\
    ={}& \eta (\mu_{v_i \rightarrow g_i}).
\end{align}
\end{subequations}
By noting the monotonicity of the logarithm function, we arrive at \eqref{prop1eq}.
\qed

\section{Proof of Proposition \ref{prop2}}\label{apdixc}
With $q(\boldsymbol{x})$ in \eqref{tildeqx}, we obtain
\begin{align}
    \mathcal{L}(v_i,q(\boldsymbol{x})) ={}& \int_{\boldsymbol{x}}q(\boldsymbol{x})\ln \frac{p(\boldsymbol{y},\boldsymbol{x}|\boldsymbol{v})|_{\mathcal{S}_{\backslash i}  }}{q(\boldsymbol{x})} \notag \\
    ={}& \int_{\boldsymbol{x}} p(\boldsymbol{x}|\boldsymbol{y},\boldsymbol{v})|_{\mathcal{S}  } \ln p(\boldsymbol{x}|\boldsymbol{v})|_{\mathcal{S}_{\backslash i}  } - \int_{\boldsymbol{x}} q(\boldsymbol{x}) \ln q(\boldsymbol{x}) \notag \\
    & + \int_{\boldsymbol{x}}  p(\boldsymbol{x}|\boldsymbol{y},\boldsymbol{v})|_{\mathcal{S}  } \ln p(\boldsymbol{y} | \boldsymbol{x}) \notag \\
    ={}& Q(v_i) + C_1,
\end{align}
where $C_1$ is a constant independent of $v_i$ and
\begin{equation}\label{Q}
    Q(v_i) = \int_{\boldsymbol{x}} p(\boldsymbol{x}|\boldsymbol{y},\boldsymbol{v})|_{\mathcal{S}} \ln p(\boldsymbol{x}|\boldsymbol{v})|_{\mathcal{S}_{\backslash i} }.
\end{equation}
Thus \eqref{L} can be recast as
\begin{equation}
    \mu_{g_i \rightarrow v_i} = \argmax_{v_i} Q(v_i).
\end{equation}
Plugging $p(\boldsymbol{x}|\boldsymbol{v})$ in \eqref{pxivi} into \eqref{Q} leads to 
\begin{equation}\label{Q2}
    Q(v_i) = -\ln v_i - \frac{1}{x_i} \int_{\boldsymbol{x}} p(\boldsymbol{x}|\boldsymbol{y},\boldsymbol{v})|_{\mathcal{S}} |x_i|^2 + C_2
\end{equation}
where $C_2$ is another constant independent of $v_i$. Further we notice
\begin{equation}
    \int_{\boldsymbol{x}} p(\boldsymbol{x}|\boldsymbol{y},\boldsymbol{v})|_{\mathcal{S}}  |x_i|^2 = |m_i|^2 + \phi_{i,i}
\end{equation}
where $m_i$ denotes the $i$-th entry of $\boldsymbol{m}$ in \eqref{m}, and $\phi_{i,i}$ denotes the $i$-th diagonal element of the covariance matrix $\boldsymbol{\Phi}$ in \eqref{phi}. Taking the derivative of \eqref{Q2} with respect to $v_i$ and setting the result to zero, we obtain \eqref{alphaupdate}.
\qed

\ifCLASSOPTIONcaptionsoff
  \newpage
\fi

\bibliographystyle{IEEEtran}
\bibliography{VSPSBLabbr2}

\end{document}